\documentclass[DIV13,11pt]{scrartcl}
\pdfoutput=1
\usepackage[utf8]{inputenc}
\usepackage{hyperref}
\usepackage{graphicx}
\usepackage{graphics}
\usepackage{dsfont}
\usepackage{epsfig}
\usepackage{amsmath,amssymb,amsthm,amscd}
\usepackage[sort&compress,square,numbers]{natbib} 
\usepackage{float}
\restylefloat{table}

\def\Mgn[#1]#2{{\overline{\cal M}_{#1,#2}}}
\def\pqs[#1,#2]{{\footnotesize{$\left[\begin{array}{c} #1\\#2  \end{array}\right]$}}} 
\def\pqsu[#1,#2]{\left[\begin{array}{c} #1\\#2  \end{array}\right]} 
\def\pqssu[#1,#2]{{\footnotesize{\left[\begin{array}{c} #1\\#2  \end{array}\right]}}} 
\def\pqh[#1,#2]{{\footnotesize{$\left[\begin{array}{c} #1\\#2  \end{array}\right]$}}} 
\def\pqhu[#1,#2]{\left[\begin{array}{c} #1\\#2  \end{array}\right]} 
\newcommand{\ba}{\begin{eqnarray*}}
\newcommand{\ea}{\end{eqnarray*}}
\newcommand{\ban}{\begin{eqnarray}}
\newcommand{\ean}{\end{eqnarray}}
\newcommand{\be}{\begin{equation}}
\newcommand{\ee}{\end{equation}}
\newcommand{\ben}{\begin{equation}}
\newcommand{\een}{\end{equation}}
\numberwithin{equation}{section}

\newcommand{\IZ}{\mathbb{Z}}
\newcommand{\IC}{\mathbb{C}}
\newcommand{\IP}{\mathbb{P}}

\newcommand{\IR}{\mathbb{R}}

\numberwithin{equation}{section}

\newcommand{\CB}{{\cal B}}
\newcommand{\CC}{{\cal C}}

\newcommand{\CF}{{\cal F}}

\newcommand{\CM}{{\cal M}}
\newcommand{\CN}{{\cal N}}
\newcommand{\CO}{{\cal O}}

\newcommand{\CW}{{\cal W}}

\def\IZ{{\mathbb Z}}
\def\IR{{\mathbb R}}
\def\IC{{\mathbb C}}
\def\IP{{\mathbb P}}

\def\IF{{\mathbb F}}

\newcommand{\ri}{{\rm i}}
\newcommand{\rd}{{\rm d}}

\newcommand{\Li}{\mathop{\rm Li}\nolimits}

\newcommand{\ts}{\thinspace}

\newcommand{\CDD}{{\cal D}}

\begin{document}
\begin{titlepage}
{}~ \hfill\vbox{ \hbox{} }\break

\rightline{
USTC-ICTS-14-03}
\rightline{
Bonn-TH-2014-02}

\vskip 3 cm

\centerline{\Large \bf  Quantum geometry of del Pezzo surfaces }  \vskip 0.5 cm
\centerline{\Large \bf  in the Nekrasov-Shatashvili limit}   \vskip 0.5 cm

\renewcommand{\thefootnote}{\fnsymbol{footnote}}
\vskip 30pt \centerline{ {\large \rm Min-xin Huang\footnote{minxin@ustc.edu.cn}, 
Albrecht Klemm\footnote{aklemm@th.physik.uni-bonn.de},
Jonas Reuter\footnote{jreuter@th.physik.uni-bonn.de}
and Marc Schiereck\footnote{marc@th.physik.uni-bonn.de}
} } \vskip .5cm \vskip 20pt

\begin{center}
{$^*$Interdisciplinary Center for Theoretical Study, University of Science and \\ \vskip 0.2 cm Technology of China,\  Hefei, Anhui 230026, China}\\ [3 mm]
{$^{\dagger\ddagger\S}$Bethe Center for Theoretical Physics and $^\dagger$Hausdorff Center for Mathematics,}\\ 
{Universit\"at Bonn, \  D-53115 Bonn}\\ [3 mm]
\end{center}

\setcounter{footnote}{0}
\renewcommand{\thefootnote}{\arabic{footnote}}
\vskip 60pt

\begin{abstract}
We use mirror symmetry, quantum geometry  
and modularity properties of elliptic curves
to calculate the refined free energies
in the Nekrasov-Shatashvili limit on non-compact toric Calabi-Yau manifolds, 
based on del Pezzo surfaces.
Quantum geometry here is to be understood as a quantum
deformed version of rigid special geometry, which has
its origin in the quantum mechanical behaviour of branes
in the topological string B-model.
We will argue that, in the Seiberg-Witten picture,
only the Coulomb parameters lead to quantum corrections,
while the mass parameters remain uncorrected.
In certain cases we will also compute the expansion of
the free energies at the orbifold point and the conifold locus.
We will compute the quantum corrections order by order
on $\hbar$, by deriving 
second order differential operators, which act on
the classical periods.
\end{abstract}

\end{titlepage}
\vfill \eject


\newpage

\baselineskip=16pt    

\tableofcontents

\section{Introduction}

The idea of quantizing geometrical structures 
originated in topological string theory from an interpretation 
of background independence~\cite{Witten:1993ed} of the 
string partition function $Z$. It turned out that the 
concept of viewing the topological string partition function $Z$ 
as a wave function on the configuration space of 
complex structures of the target space~\cite{Witten:1993ed} 
plays a central r\^ole in black hole physics~\cite{Ooguri:2004zv}, 
in calculating world sheet-~\cite{Aganagic:2006wq} as well 
as space time instanton expansions~\cite{Nekrasov:2002qd,
Nekrasov:2009rc}, for large N-dualities and in holographic
applications~\cite{Marino:2009jd}. In context of topological 
string theory on Calabi-Yau  geometries leading to $\CN=2$ theories in 4d one can view 
this as a quantization of special geometry with the genus 
expansion parameter $g_s^2$ playing the r\^ole of $\hbar$. 

After Nekrasov~\cite{Nekrasov:2002qd} introduced the $\Omega$-background 
with two deformation parameters $\epsilon_1$ and $\epsilon_2$ 
to regularize the moduli space of instantons in $\CN=2$  Super-Yang-Mills 
theories, it became quickly clear~\cite{Nekrasov:2002qd}, how to 
interpret these two parameters in the topological string partition 
function on local Calabi-Yau spaces, which are related to the gauge 
theories by geometric engineering. In fact the work 
of~\cite{Gopakumar:1998jq}\cite{Katz:1999xq} anticipated 
a geometrical interpretation of the latter in terms of a refined counting 
of BPS states corresponding to $D0-D2$ branes the large volume limit of 
rigid $\CN = 2$ theories in four dimensions.  The multiplicities 
$N^\beta_{j_L,j_R}\in \mathbb{N}$ of the refined BPS states lift the 
degeneracy of the $j_R$ spin-multiplets that is present in the 
corresponding BPS index $n^\beta_g\in \mathbb{Z}$ of the topological 
string, which correspond to the specialization $i g_s=\epsilon_1=-\epsilon_2$. 
A mathematical definition of the refinement of cohomology of the moduli 
space of the BPS configurations, was recently given~\cite{Choi:2012jz} 
starting with the moduli space of Pandharipande-Thomas invariants.  

In another development it was pointed out in~\cite{Aganagic:2011mi} 
based on the earlier work~\cite{Aganagic:2003qj} that the \emph{Nekrasov-Shatashvili} 
limit $\epsilon_1=0$~\cite{Nekrasov:2009rc} provides an even simpler 
quantization description of special geometry in which the r\^ole of $\hbar$ 
is now played by $\epsilon_1$ (or equivalently $\epsilon_2$) and the r\^ole of the configuration space 
is played by the moduli space of a brane, which is identified with the 
B-model curve itself.       

The free energies $F=\log(Z)$ of the topological string at 
large radius in terms of the  BPS numbers $N^\beta_{j_L j_R}$ are 
obtained by a Schwinger-Loop calculation~\cite{Iqbal:2007ii,Huang:2010kf} and read 
\begin{equation}
\begin{array}{rl}
F^{hol}(\epsilon_1,\epsilon_2,t)&=
\displaystyle{\sum_{ \genfrac{}{}{0pt}{}{j_L,j_R=0}{k=1} }^\infty \sum_{\beta\in H_2(M,\mathbb{Z})} (-1)^{2(j_L+j_R)} \frac{N^\beta_{j_L j_R}}{k}
\frac{\displaystyle{\sum_{m_L=-j_L}^{j_L}} q_L^{k m_L}}{2\sinh\left( \frac{k \epsilon_1}{2}\right)} \frac{\displaystyle{\sum_{m_R=-j_R}^{j_R}} q_R^{k m_R}}
{2\sinh\left( \frac{k \epsilon_2}{2}\right)}e^{-k\, \beta \cdot t}}\ .
\end{array}
\label{schwingerloope1e2_int}
\end{equation}
where 
\begin{equation}
 q_L = \exp{ \frac{1}{2}(\epsilon_1 - \epsilon_2) } \quad\text{and}\quad{}q_R = \exp{ \frac{1}{2}(\epsilon_1 + \epsilon_2) }\,.
\end{equation}
This expression admits an expansion in $\epsilon_1, \epsilon_2$ in the following way
\begin{equation}
F(\epsilon_1,\epsilon_2,t)={\rm log}(Z) = \sum_{n,g=0}^{\infty} (\epsilon_1+\epsilon_2)^{2n}
(\epsilon_1\epsilon_2)^{g-1} F^{(n,g)}(t)\, .
\label{eqn:F_e1e2exp}
\end{equation}
This defines the refinement of the free energies as a two parameter deformation
of the unrefined topological string. The usual genus expansion of
the unrefined string is just encoded in $F^{0,g}$, which we obtain by
setting $\epsilon_1 = - \epsilon_2$.

Techniques to compute this instanton series already exist.
Starting with the mathematical definition one can do now a direct 
localisation calculation~\cite{Choi:2012jz}. Alternatively  one can use the 
refined topological vertex~\cite{Iqbal:2007ii}, or the holomorphic anomaly equation, 
which has been generalized for the use in the refined case 
in~\cite{Huang:2010kf,Krefl:2010fm,Krefl:2010jb}.

In this paper we will consider the \emph{Nekrasov-Shatashvili} limit, i.e. we set 
one of the deformation parameters, say $\epsilon_1$, in \eqref{eqn:F_e1e2exp}   
to zero and expand in the remaining one $\epsilon_2 = \hbar$. 
In \cite{Nekrasov:2009rc} Nekrasov and Shatashvili conjectured that 
this limit leads to a description of the presented setup 
as a quantum integrable system. Looking at the expansion given in
\eqref{eqn:F_e1e2exp}, we see that the Nekrasov-Shatashvili limit 
is encoded in the terms $F^{(n,0)}$ of the full free energy.

We base our calculation on the results of \cite{Aganagic:2011mi}, where branes
were studied in the context of refined topological strings. 
Branes probe the geometry
in a quantum mechanical way, which was analyzed in \cite{Aganagic:2003qj} for
the B-model on Calabi-Yau geometries given by
\begin{equation}
  u v = H(x, p; z) \,,   \label{eqn:geom_RS}
\end{equation}
where $H(x,p; z) = 0$ defines a Riemann surface.
The wave function $\Psi(x)$ which describes such branes
satisfies the operator equation
\begin{equation}
  \hat{H} \Psi(x) = 0 \label{eqn:RS_op_eq} 
\end{equation}
where $\hat{H}$ is defined via the position space
representation of $p$ if interpreted as the momentum 
\begin{equation}
  \hat{H} = H(x, \ri \hbar \partial_x )\,.  \label{eqn:RS_op} 
\end{equation}
$\hat{H}$ reduces to the Riemann surface $\eqref{eqn:geom_RS}$ in the semiclassical limit. 
In case of the unrefined topological string one obtains further
corrections in $g_s$ to $\hat{H}$, so that the relation \eqref{eqn:RS_op_eq} is only
true to leading order.

In the refined topological string two types of branes exist which correspond to $M5$ branes wrapping different cycles in the M-theory lift. 
The way these branes probe the geometry is a key ingredient for 
deriving the results in this article.

A refinement of the topological B-model in terms of a matrix model
has been conjectured in \cite{Dijkgraaf:2009pc}. This refinement, based
on the matrix model description of the topological B-model, amounts to
deforming the Vandermonde determinant in the measure by a power 
of $\beta = - \frac{\epsilon_1}{\epsilon_2}$.
By virtue of this matrix model the time dependent Schr\"odinger equation
\begin{equation}
  \hat{H} \Psi = \epsilon_1 \epsilon_2 \sum f_I(t) \frac{\partial \Psi}{\partial t_I} \label{eqn:td_sglgint}
\end{equation}
can be derived. $\hbar$ is either identified with $\epsilon_1$ or with $\epsilon_2$,
depending on which brane the wavefunction describes.

In the Nekrasov-Shatashvili limit we have $g_s \rightarrow 0$, therefore
this picture simplifies immensely and relation
\eqref{eqn:RS_op} is true up to normal ordering ambiguities.
This can be seen from the Schr\"odinger equation \eqref{eqn:td_sglgint}

From the result of moving the branes around cycles of the geometry one can
deduce that the free energy in the Nekrasov-Shatashvili limit can
be computed via the relation of special geometry between A- and B-cycles.
We will have to introduce a deformed differential over which periods
of these cycles are computed.

This setup is conjectured to be true generally and in this paper we
want to check it for more general geometries. 
Furthermore we aim to clear up the technical implementations of
this computation. 
This means we want to identify the right parameters of the models
and compute the free energies in a more concise way.
In case of local Calabi-Yau geometries, we find two different kind of moduli.
These are normalizable and non-normalizable moduli. In order to
successfully compute the free energies, we have to keep this difference in
mind, especially because the non-normalizable moduli will not obtain
any quantum corrections.

In the context of Seiberg-Witten theory an interpretation of these
distinction exists in the sense that the normalizable moduli
are related to the Coulomb parameters 
while the non-normalizable are identified as mass parameters of the gauge theory,
which appear as residues of the meromorphic differential defined on the
Seiberg-Witten curve.

We use the relations introduced in \cite{Aganagic:2011mi}  and 
apply them to the case of mirror duals of toric varieties.
In order to compute higher order corrections to the quantum
deformed meromorphic differential, we derive certain
differential operators of order two.
Based on \cite{Mironov:2009uv} this method has been used in \cite{Aganagic:2011mi} 
for the  cubic matrix model and it has been
applied in \cite{Huang:2012kn} to the case of toric geometries.
We find that these operators act only on the normalizable moduli.
There are some advantages in using these operators, one is that
we are able to compute the free energies in different regions of the moduli space,
another is that we do not have to actually solve the period integrals.
This method of computing the higher order corrections also clears up 
their structure. Namely, the mass parameters will not obtain any quantum
corrections, while the periods, do.

We will compute the free energies in the Nekrasov-Shatashvili limit of
the topological string on local Calabi-Yau geometries with del Pezzo
surfaces or mass deformations thereof as the base.
For the local $\IP^2$ we also compute it at
different points in moduli space namely, not only the large radius point,
but also at the orbifold point and the conifold locus. For local $\IF_0$ we
also will not only solve the model at large radius, but also
at the orbifold point.

In section \ref{sec:geom_set} we will provide an overview of the
gemetric structures we are using. In \ref{sec:quantum_periods}
we introduce the Nekrasov-Shatashvili limit and motivate
a quantum special geometry, which we use to finally solve
the topological string in the Nekrasov-Shatashvili limit in
section \ref{sec:computations}.
%
%
%
%
%
%
\section{Geometric setup}\label{sec:geom_set}
\subsection{Branes and Riemann surfaces}
We want to strengthen the conjecture made in \cite{Aganagic:2011mi}
and clear up some technical details of this computation along the way.
Let us therefore briefly review the geometric setup which we
consider here.

Similarly to computations that were performed in \cite{Huang:2013yta}
we want to compute the instanton series of the topological string A-model 
on non-compact Calabi-Yau spaces $X$, which are given as the total space
of the fibration of the anti-canonical line bundle
\begin{equation}
  \CO(- K_B ) \rightarrow B
\end{equation}
over a Fano variety $B$. By the adjunction formula this defines a non-compact 
Calabi-Yau $d$-fold for $(d-1)$-dimensional Fano varieties. \emph{Del Pezzo}
surfaces are two-dimensional smooth Fano manifolds and they enjoy a finite
classification. These consists of $\IP^2$ and blow-ups of $\IP^2$ in up to
$n=8$ points, called $\CB_n$, as well as $\IP^1 \times \IP^1$.

As a result of mirror symmetry we are able to compute the amplitudes
in the topological string B-model, where the considered geometry is given by
\begin{equation}
  u v = H(e^p, e^x; z_I)
  \label{eqn:uvH}
\end{equation}
with $u,v \in \IC$, $e^p,e^x \in \IC^*$ and $z_I$ are complex strucure moduli.
Furthermore $H(e^p,e^x; z_I) = 0$ is the defining equation of a Riemann surface.

The analysis in the following relies heavily on the insertion of
branes into the geometry and their behaviour when moved around
cycles.
Let us continue along the lines of \cite{Aganagic:2003qj} with the description
of the influence branes have if we insert them into this geometry.
In particular let us consider 2-branes. 
If we fix a point $(p_0, x_0)$ on the $(p,x)$-plane
these branes will fill the subspace of fixed $p_0$, $x_0$, 
where $u$ and $v$ are restriced by 
\begin{equation}
  u v = H(p_0, x_0) .
\end{equation}
The class of branes in which we are interested, corresponds to fixing $(p_0, x_0)$ 
in a manner so that they lie on the Riemann surface, i.\,e.
\begin{equation}
  H(p_0, x_0) = 0 \,.
\end{equation}
By fixing the position of the brane like this, the moduli space of the brane
is given by the set of admissible points, meaning it can be identified with the
Riemann surface itself.

Following from an analysis of the worldvolume theory of these branes, one can argue that the
two coordiantes $x$ and $p$ have to be noncommutative.
Namely, this means that they fulfill the commutator relation
\begin{equation}
  [x, p] = g_s \,,
\end{equation}
where $g_s$ is the coupling constant of the topological string, which
takes the role of the Planck constant.

The leading order part of such a brane's partition function is given by
\begin{equation}
  \Psi_\text{cl.}(x) = \exp \left( \frac{1}{g_s} \int^x p(y) \rd y \right) . \label{eqn:WKB_lead_unref}
\end{equation}
This looks a lot like the first order term of a WKB approximation if we would identify
$H(x,p)$ with the Hamiltonian of the quantum system. 
All of this suggests that $\Psi(x)$ is a wave-function for the quantum Hamiltonian $H$.
As a result, we are expecting a relation of the form
\begin{equation}
  \hat{H}(x, p) \Psi(x) = 0\,,
\end{equation}
which can be considered as $H(x, p) = 0$ written as a condition
on operators.
Unfortunately it is generally not possible to derive this Hamiltonian, because
we do not have control over the higher order $g_s$-corrections to it.
But this is the story for the unrefined case. In the Nekrasov-Shatashvili limit
of the refined topological string this problem disappears as we will show later on.
\subsubsection{Mirror symmetry for non-compact Calabi-Yau spaces}
We want to analyze toric del Pezzo surfaces and mass deformations thereof.
These kind of geometries are related to Riemann surfaces defined by equations like \eqref{eqn:uvH} via mirror symmetry.
Given the toric data of  a non-compact Calabi-Yau space, there
exists a construction which gives the defining equation for
the Riemann surface.

The A-model geometry of a noncompact toric variety is given by
a quotient
\begin{equation}
  M = ( \IC^{k+3} \setminus \mathcal{SR} ) / G,
\end{equation}
where $G = (\IC^*)^k$ and $\mathcal{SR}$ is the Stanley-Reisner ideal. The group $G$ acts on the homogeneous coordinates $x_i$
via
\begin{equation}
  x_i \rightarrow \lambda_\alpha^{ l_i^\alpha } x_i
\end{equation}
where $\alpha = 1,\ldots, k$ and $\lambda_\alpha \in \IC^*$, $l_i^\alpha \in \IZ$.
The Stanley-Reisner ideal needs to be chosen in a way that the variety $M$ exists. 
The toric variety $M$ can also be viewed as the vacuum field configuration of a 2d abelian (2,2) gauged linear $\sigma$-model.
In this picture the coordinates $x_i \in \IC^*$ are the
vacuum expectation values of chiral fields. These fields transform
as
\begin{equation}
   x_i \rightarrow e^{\ri l_i^\alpha \epsilon_\alpha} x_i 
\end{equation} 
under the gauge group $U(1)^k$,
where again $l_i^\alpha \in \IZ$ and $\alpha = 1, \ldots, k$, 
while $\epsilon_\alpha \in \IR$.

The vacuum field configurations are the equivalence classes under the gauge group,
which fulfill the $D$-term constraints
\begin{equation}
  D^\alpha = \sum_{i = 1}^{k + 3} l_i^\alpha 
|x_i|^2 = r^\alpha, \quad \alpha = 1,\ldots, k
  \label{eqn:D_term_const}
\end{equation}
where the $r^\alpha$ are the K\"ahler parameters. In string theory $r^\alpha$ is complexified
to $T^\alpha = r^\alpha + \ri \theta^\alpha$.
The Calabi-Yau condition $c_1(TM) = 0$ is equivalent to the anomaly condition
\begin{equation}
  \sum_{i = 1}^{k + 3} l_i^\alpha = 0, \quad \alpha = 1,\ldots, k \, .
\end{equation}
Looking at \eqref{eqn:D_term_const}, we see that negative entries in the $l$-vectors
lead to noncompact directions in $M$. 

But we are going to do computations in the topological string B-model
defined on the mirror $W$ of $M$.  We will now describe briefly how $W$
will be constructed. 
Let us define $x_i := e^{y_i} \in C^{*}$, where $i = 1, \cdots, k+3$
are homogeneous coordinates.
Using the charge vectors $l^\alpha$, we define coordinates
$z_\alpha$ by setting
\begin{equation}
 z_\alpha = \prod_{i = 1}^{k + 3} x_i^{l_i^\alpha},\quad \alpha = 1,\ldots,k \,.
  \label{eqn:batcoord}
\end{equation}
These coordinates are called \emph{Batyrev coordinates} and are chosen so that 
$z_\alpha = 0$ at the large complex structure point.
In terms of the homogeneous coordinates a Riemann surface can
be defined by writing
\begin{equation}
  H = \sum_{i=1}^{k+3} x_i\,.
  \label{eqn:rs_bat}
\end{equation}
Using \eqref{eqn:batcoord} to eliminate the $x_i$ and setting one $x_i = 1$ ,
we are able to parameterize the Riemann surface \eqref{eqn:rs_bat} via two variables,
which we call
$X= \exp(x)$ and $P = \exp(p)$. 
Finally, the mirror dual $W$ is given by the equation
\begin{equation}
  u v = H(e^x, e^p; z_I) \quad I = 1, \ldots, k\,.
\label{eqn:mirrorcurve}
\end{equation}
%
%
%
%
%
%
%
%
\section{The refinement}
\label{sec:quantum_periods}
This was the story for the unrefined case, but we actually are interested
in the refined topological string. Let us therefore introduce the
relevant changes that occur when we consider the refinement of the
topological string.
According to \cite{Dijkgraaf:2006um}, the partition function of the topological
A-model on a Calabi-Yau X is equal to the partition function of M-theory
on the space
\begin{equation}
  X \times TN \times S^1 \label{eqn:mth_geom}
\end{equation}
where $TN$ is a Taub-NUT space, with coordinates $z_1, z_2$.
The $TN$ is fibered over the $S^1$ so that, when going
around the circle, the coordinates $z_1$ and $z_2$ are twisted by
\begin{equation}
  z_1 \rightarrow e^{i \epsilon_1}z_1 \quad\text{and}\quad z_2 \rightarrow e^{i \epsilon_2} z_2 .
\end{equation}
This introduces two parameters $\epsilon_1$ and $\epsilon_2$ and 
unless $\epsilon_1 = - \epsilon_2$ supersymmetry is broken. 
But if the Calabi-Yau is non-compact we are able to relax this condition,
because an addtional $U(1)_R$-symmetry, acting on $X$, exists.

General deformations in $\epsilon_1$ and $\epsilon_2$ break the symmetry between $z_1$ and $z_2$
of the Taub-NUT space in \eqref{eqn:mth_geom}. As a result we find two types of branes
in the refinement of the topological string. 
In the M-theory setup the difference is given by
the cigar subspaces $\IC \times S^1$ in $TN \times S^1$ of \eqref{eqn:mth_geom},
which the M5-brane wraps. 

The classical partition function of an $\epsilon_i$-brane is now given by
\begin{equation}
  \Psi_{i, \text{cl.}(x)} = \exp \left( \frac{1}{\epsilon_i} W(x) \right)\,, \label{eqn:WKB_lead_ref}
\end{equation}
where $W(x)$ is the superpotential of the $\CN = (2,2)$, $d=2$ world-volume theory
on the brane and which is identified with the $p$-variable in \eqref{eqn:mirrorcurve} as
\begin{equation}
 W(x) = - \int^x p(y) \rd y \, .
\end{equation}
This is quite similar to \eqref{eqn:WKB_lead_unref} and still looks like the leading
order contribution of a WKB expansion where only the coupling changed.

This suggests that the $\epsilon_{1/2}$-branes themselves also behave like quantum objects
and if we have again say an $\epsilon_1$-brane with only one point lying on the Riemann surface parameterized by $(p, x)$ then the two coordinates are 
again noncommutative, i.\,e.
\begin{equation}
  [x, p] = \epsilon_1 = \hbar\,.
\end{equation}
We will show later that the free energy of the refined topological string can be extracted from a brane-wave function like this
in a limit where we send either one of the $\epsilon$-parameters to zero. The limit
of $\epsilon_i$ to zero means that one of the branes of the system decouples.
In the next section we will describe the relevant limit.
%
%
%
%
%
%
\subsection{The Nekrasov-Shatashvili limit}
In \cite{Nekrasov:2009rc} the limit where one of the deformation parameters is set to zero was introduced.
The free energy in this so called \emph{Nekrasov-Shatashvili} limit is defined by
\begin{equation}
  \CW(\hbar) = \lim_{\epsilon_2 \rightarrow 0} \epsilon_1 \epsilon_2 F .  
  \label{eqn:NS_limit}
\end{equation}
where $\CW$ is the called the \emph{twisted superpotential}.
This $\CW$ can be expanded in $\hbar$ like
\begin{equation}
  \CW(\hbar) = \sum_{n=0} \hbar^{2n} \CW^{(n)}
\end{equation}
where the $\CW^{(i)}$ can be identified like
\begin{equation}
  \CW^{(i)} = F^{(i,0)}
\end{equation}
with the free energy in the expansion \eqref{eqn:F_e1e2exp}.

Because we are only computing amplitudes in this limit, we present a convenient
definition of the instanton numbers, tailored for
usage in this limit.
We define the parameters
\begin{equation}
  \epsilon_L = \frac{\epsilon_1 - \epsilon_2}{2},\quad  \epsilon_R = \frac{\epsilon_1 - \epsilon_2}{2}
\end{equation}
and accordingly
\begin{equation}
  q_{1,2} = e^{\epsilon_{1,2}},\quad q_{L,R} = e^{\epsilon_{L,R}} .
\end{equation}
Using this definition the free energy at large radius has the following
expansion 
\begin{equation}
\begin{array}{rl}
F^{hol}(\epsilon_1,\epsilon_2,t)&=
\displaystyle{\sum_{ \genfrac{}{}{0pt}{}{j_L,j_R=0}{k=1} }^\infty \sum_{\beta\in H_2(M,\mathbb{Z})} (-1)^{2(j_L+j_R)} \frac{N^\beta_{j_L j_R}}{k}
\frac{\displaystyle{\sum_{m_L=-j_L}^{j_L}} q_L^{k m_L}}{2\sinh\left( \frac{k \epsilon_1}{2}\right)} \frac{\displaystyle{\sum_{m_R=-j_R}^{j_R}} q_R^{k m_R}}
{2\sinh\left( \frac{k \epsilon_2}{2}\right)}e^{-k\, \beta \cdot t}}
\end{array}
\label{schwingerloope1e2}
\end{equation}
in terms of BPS numbers $N^\beta_{j_L j_R}$.

By a change of basis of the spin representations  
\begin{equation}
\sum_{g_L,g_R} n^\beta_{g_L,g_R} I_L^{g_L}\otimes I_R^{g_R}= \sum_{j_L,j_R}
N^\beta_{j_L,j_R} \left[\frac{j_L}{2}\right]_L \otimes \left[\frac{j_R}{2}\right]_R
\label{n_basis}
\end{equation}
we introduce the instanton numbers $n^\beta_{g_R,g_L}$, which are more convenient to
extract from our computations.
With the sum over the spin states given by the expression
\begin{equation}
  \sum_{m=-j}^{j} q^{k m} = \frac{q^{ j + \frac{k}{2} } - q^{ -j - \frac{k}{2} } }{q^{\frac{k}{2}} - q^{-\frac{k}{2}} } = \chi(q^\frac{k}{2})
\end{equation}
we write down the relation between $ N^\beta_{j_L j_R}$ and the numbers  $n^\beta_{g_R,g_L}$
defined in \eqref{n_basis} explicitly \cite{Huang:2010kf,Hatsuda:2013oxa}
\begin{equation}
	\sum_{j_L, j_R} (-1)^{2(j_L+j_R)} N^\beta_{j_L j_R}
	\chi(q_L^\frac{k}{2}) \chi(q_R^\frac{k}{2}) 
	= \sum_{g_L, g_R} n^\beta_{g_L,g_R}(q_L^{\frac{1}{2}} - q_L^{-\frac{1}{2}} )^{2 g_L} (q_R^{\frac{1}{2}} - q_R^{-\frac{1}{2}})^{2 g_R}\ .
\end{equation}
Since we do not consider the full refined topological string we want to see how this expansion looks like in the Nekrasov-Shatashvili limit. 
Writing \eqref{schwingerloope1e2} in terms of $n^\beta_{g_L,g_R}$ and 
taking the Nekrasov-Shatashvili limit \eqref{eqn:NS_limit}, we find
\begin{equation}
\begin{array}{rl}
	\CW(\hbar, t)&= \hbar
		\displaystyle{\sum_{ \genfrac{}{}{0pt}{}{g=0}{k=1} }^\infty 
		\sum_{\beta\in H_2(M,\mathbb{Z})}
		\frac{\hat{n}^\beta_{g}}{k^2}
		\frac{ (q^{\frac{k}{4}} - q^{-\frac{k}{4}} )^{2 g} }{2\sinh\left( \frac{k \hbar }{2}\right)} 
		e^{-k\, \beta \cdot t}}
\end{array}
\label{eqn:Wexp}
\end{equation}
where $\hbar=\epsilon_1$ and
\begin{equation}
   \hat{n}^\beta_{g} = \sum_{g_L + g_R = g} n^\beta_{g_L,g_R} \, .
\end{equation}
%
%
%
%
%
%
\subsection{Schr\"odinger equation from the  \texorpdfstring{$\beta$}{beta}-ensemble}
In \cite{Aganagic:2003qj} the authors described the behavior of branes
by analyzing the relevant insertions into the matrix model description of
the topological string B-model.
In \cite{Dijkgraaf:2009pc} a conjecture has been made about a matrix model description of
the refined topological B-model, which we now want to use as described in \cite{Aganagic:2011mi} to derive a Schr\"odinger equation for the brane-wavefunction
of an $\epsilon_1$ or $\epsilon_2$-brane.
This matrix model takes the form of a deformation
of the usual matrix model, describing the unrefined topological string where the usual Vandermonde-determinant is not taken to the second power anymore, but to the power $2 \beta$ where 
\begin{equation}
  \beta = -\frac{\epsilon_1}{\epsilon_2}.
\end{equation}
This clearly has the unrefined case as its limit, when $\epsilon_1 \rightarrow - \epsilon_2$. 
Matrix models of this type are called $\beta$-ensembles.

The partition function of this matrix model is
\begin{equation}
 Z = \int \rd^N z \prod_{i < j} (z_i - z_j)^{-2 \epsilon_1 / \epsilon_2}  e^{ - \frac{2}{\epsilon_2} \sum_i W(z_i) } .
\end{equation}
The free energy of this matrix model can be expanded in $g_s$ and $\beta$ in the following way
\begin{equation}
  F=\sum_{n,g=0} \gamma^{2 n} g_s^{2 n + 2 g -2} F_{n,g}
\end{equation}
where we defined 
\begin{equation}
  \gamma = \sqrt{\beta} - \sqrt{\beta^{-1}} \,.
\end{equation}
Here we used
\begin{equation}
  \epsilon_1 = \ri \sqrt{\beta} g_s \quad \epsilon_2 = - \ri \frac{g_s}{ \sqrt{\beta} }\,.
\end{equation}
This gives the expansion \eqref{eqn:F_e1e2exp} in terms of $\epsilon_1$ and $\epsilon_2$
if we identify
\begin{equation}
	F_{n,g} = (-1)^n F^{(n,g)}\,.
\end{equation}
Based on this matrix model description the following equation for brane wave-functions has been derived in \cite{Aganagic:2011mi}
\begin{equation}
  \left( - \epsilon_\alpha^2 \frac{\partial^2}{\partial x^2} + W'(x)^2 + f(x) + g_s^2 \sum_{n=0}^g x^n \partial_{(n)} \right) \Psi_\alpha(x) = 0\,.
\end{equation}
Now let us take the Nekrasov-Shatashvili limit. Here we consider the case
\begin{equation}
  \hbar = \epsilon_1,\quad\text{and}\quad \epsilon_2 \rightarrow 0 .
\end{equation}
Due to the identity $g_s^2 = - \epsilon_1 \epsilon_2$, the term containing $g_s^2$ vanishes leaving us with a time independent Schr\"odinger equation. 
(For a more detailed explanation
of what is meant by this see \cite{Aganagic:2011mi}.)
We are left with a time-independent Schr\"odinger equation for the
$\epsilon_1$-brane, where the $\epsilon_2$-brane decouples.

If we now interpret
\begin{equation}
  \ri \hbar \frac{\partial}{\partial x} = \hat{p}
\end{equation}
as the position-space representation of the momentum operator $\hat{p}$
this yields the form
\begin{equation}
  ( \hat{p}^2 + (W'(x))^2 + f(x) ) \Psi(x) = 0\, ,
\end{equation}

where $\Psi(x) = \Psi_2(x)$ is the brane partition function of the brane which does
not decouple when taking the Nekrasov-Shatashvili limit.

In the limit $\hbar \rightarrow 0$ this equation becomes classical and we are left
with the defining equation of the Riemann surface
\begin{equation}
  p^2 + W'(x)^2 + f(x) = 0 \, .
\end{equation}

Having such a matrix model description, we are able to describe the effect
the insertion of branes into the geometry has. In the unrefined case, the 
meromorphic differential $\lambda$ acquires a pole with residue $g_s$
at the point the brane was inserted. Therefore by
going around the position $x_0$ of this brane, we pick up
\begin{equation}
  \oint_{x_0} \lambda = g_s \,.
\end{equation}
This behavior is captured by the Kodaira-Spencer scalar field $\phi$ on $\Sigma$ by the relation
\begin{equation}
  \delta \lambda = \partial \phi\,.
\end{equation}
Via bosonization we can relate this to the insertion of the brane insertion operator
\begin{equation}
  \psi(x) = e^{\phi / g_s}
\end{equation}
which is a fermion.
In terms of periods this means
\begin{equation}
 \oint_{x_0} \partial \phi \psi(x_0) = g_s \psi(x_0) \, .
\end{equation}
In analogy to \eqref{eqn:WKB_lead_ref} we define the brane insertion operator
in the refined case as
\begin{equation}
  \psi_\alpha(x) = \exp( \phi(x) / \epsilon_\alpha )\quad \alpha = 1,2
\end{equation}
and the Riemann surface is deformed in a similar manner by an
$\epsilon_i$-brane inserted at the point $x_0$
\begin{equation}
  \oint_{x_0} \partial \phi \psi_i(x_0) = \frac{g_s^2}{\epsilon_i} \psi_i (x_0)\,.
\end{equation}
\subsection{Special geometry}\label{sec:special_geometry}
Up to now we learned that the branes we are considering act like quantum theoretic objects.
In order to make use of this, we derived Schr\"odinger equations for the wave functions of
$\epsilon_1$- and $\epsilon_2$-branes, respectively. However we are actually interested
in deriving free energies. 

This will be achieved by a deformed version of special geometry.
But to make things more clear let us put this into a more general context and give a very short introduction to special geometry.
Via special geometry we are able to derive the genus zero contribution
of the full free energy which we will call the prepotential.

We start with introducing the \emph{periods} of the holomorphic three-form $\Omega$
of a Calabi-Yau threefold $X$. The first step is choosing a basis of three
cycles $A^I$ and $B_J$, where $I,J = 0, \ldots, h^{2,1}$, with intersection numbers
\begin{equation}
  A^I \cap B_J = - B_J \cap A^I, \quad A^I \cap A^J = B_I \cap B_J = 0 \, .
\end{equation}
The dual cohomology basis spanning $H^3(X, \IZ)$
\begin{equation}
  (\alpha_I, \beta^I ), \quad I=0,1, \cdots h^{2,1}(X) \label{eqn:symbasis3form}
\end{equation}
is given by  Poincaré duality
\begin{equation}
  \int_{A^I} \alpha_I  = \delta_I^J,\quad \int_{B_J} \beta^J = - \delta^J_I
\end{equation}
and satisfies the relations
\begin{equation}
  \int_X \alpha_I \wedge \beta^J = \delta_I^J\quad \text{and}\quad \int_X \beta^J \wedge \alpha_I = - \delta^J_I
\end{equation}
while all other combinations vanish.

Now we are able to define the \emph{periods} of the holomorphic 3-form $\Omega$ by
\begin{equation}
 X^I = \int_{A^I} \Omega, \quad \CF_I = \int_{B_I} \Omega \, .
 \label{eqn:basis_SG}
\end{equation}
These periods carry information about the complex structure deformations.
The holomorphic three-form $\Omega$, as an element of $H^3(X, \IC)$,
can be expressed in terms of the basis \eqref{eqn:symbasis3form} in the following way
\begin{equation}
  \Omega = X^I \alpha_I - \CF_I \beta^I \, .
\end{equation}
The $X^I$ can locally serve as homogeneuous coordinates of the
moduli space $\CM$.
From these we choose a nonzero coordinate, e.\,g. $X^0$ and define
\begin{equation}
  t^a = \frac{X^a}{X^0}, \quad a=1,\cdots, h^{2,1}(X) 
  \label{eqn:tcoord}
\end{equation}
which are flat coordinates for the moduli space $\CM$.
The $X_I$ and $\CF^I$ are not independent and we can derive from
the fact
\begin{equation}
  \int_X \Omega \wedge \frac{\partial}{\partial X^I} \Omega = 0,\qquad 
  \int_X \Omega \wedge \frac{\partial}{\partial X^I}\frac{\partial}{\partial X^J}  \Omega = 0
\end{equation}
that a holomorphic function $\CF$ exists, which we	 will
call the \emph{prepotential}. 
This prepotential obeys the relations
\begin{equation}
  \CF = \frac{1}{2} X^I \CF_I,\quad \CF_I = \partial_{X^I} \CF \, ,
\end{equation}
which imply that $\CF$ is homogeneous of degree two in $X^I$.
In flat coordinates we define
\begin{equation}
 \CF(X_I) = (X^0)^2 F(t^I)\,,
\end{equation}
which fulfills the relations
\begin{equation}
  F_I = \frac{\partial F}{\partial t^I}\, .
  \label{eqn:sgrel}
\end{equation}
Since we are analyzing local Calabi-Yau spaces we have to
consider \emph{rigid special geometry}.
Here we will analyze only the B-model topological string on
local Calabi-Yau threefolds $X$ which are given by the equation
\begin{equation}
 u v = H(e^x, e^p; z_I) 
\end{equation}
as we stated before in section \ref{sec:geom_set}. 
The holomorphic three-form $\Omega$ in this case is given by
\begin{equation}
  \Omega = \frac{\rd u }{u} \wedge \rd x  \wedge \rd p .
\end{equation}
The three-cycles on $X$ descend to one-cycles on the Riemann surface $\Sigma$
given by the equation
\begin{equation}
   H(e^x, e^p; z_I) = 0\,.
\end{equation}
Furthermore we find the relation that the periods of the holomorphic three form on
the full Calabi-Yau threefold descend to periods of a meromorphic one-form $\lambda$,
on only the Riemann surface $\Sigma$. This one-form is given by
\begin{equation}
  \lambda = p \rd x \,.
\end{equation}
Hence we can concentrate on the geometry of Riemann surfaces. There are $2 g$ compact
one-cycles on a genus $g$ surface. These form a basis with the elements $A^i$ and $B_i$, where $i$ runs
from $1$ to $g$. We demand their intersections to be
\begin{equation}
   A^i \cap B_j = \delta^i_j
\end{equation}
or more generally equal to $n^i_j$, with $n^i_j$ being an integer.

Having found this basis, we define the periods of the meromorphic one-form
\begin{equation}
   x^i = \oint_{A^i} \lambda, \quad p_i = \oint_{B_i} \lambda, 
   \label{eqn:RS_x_p}
\end{equation}
analogously to \eqref{eqn:basis_SG}. 
Here the $x^i$ are normalizable moduli of the Calabi-Yau manifold. 
But we are considering non-compact Calabi-Yau manifolds
and the non-compactness leads to additional non-normalizable moduli.
These are mere parameters, not actual moduli of the geometry.

The normalizable moduli are related to the Coulomb parameters in 
Seiberg-Witten theory, e.\,g. pure $SU(N)$ Seiberg-Witten theories 
have an $N-1$ Coulomb parameters, which correspond to the $g=N-1$ period 
integrals over the $A$-cycles of the genus $g$ Seiberg-Witten curve.

We already introduced the meromorphic differential $\lambda$ coming
from a reduction of the holomorphic three-form on the Riemann surface $\Sigma$.
In Seiberg-Witten theory one can have additional periods on $\Sigma$
for theories with matter. 
These periods arise because $\lambda$ has poles
in this case and the residues correspond to \emph{mass parameters}.
This explains why we need to separate two types of moduli in terms
of a physical interpretation.
%
%
%
%
%
%
\subsection{Quantum special geometry}
In \cite{Aganagic:2011mi} it was derived that the free energies of the topological string
in the Nekrasov-Shatashvili limit can be derived by taking the defining equation for the
Riemann surface and use it as the Hamiltonian of the system, which is then quantized.
In this case the non-vanishing $\epsilon$-parameter will take the role of the
Planck constant. 
Which parameter we choose does not affect the computation, so let us set
\begin{equation}
  \hbar = \epsilon_1\, .
\end{equation}
The $\epsilon_2$-parameter will be sent to zero, which amounts to the decoupling
of the $\epsilon_2$-branes.
In order to quantize this system we interpret $x$ and $p$
as canonically conjugated coordinates and lift them to operators
$\hat{x}$ and $\hat{p}$.
On these operators we impose the commutation relation
\begin{equation}
  [ \hat{x}, \hat{p} ] = \ri \hbar
\end{equation}
so that $\hat{p}$ will be
\begin{equation}
 \hat{p} = \ri \frac{\partial}{\partial x}
\end{equation}
in $x$-space. The reasoning behind this is that the $\epsilon_i$-branes
still behave like quantum mechanical objects. 

We quantize the system as described above by letting the defining equation for the Riemann surface
become the differential equation
\begin{equation}\label{eqn:HamiltonOperator}
  H(x, \ri \hbar \partial_x) \Psi(x) = 0\,.
\end{equation}
One way to solve this differential equation is the WKB method, where we use the ansatz
\begin{equation} \label{eqn:WKBansatz}
  \Psi(x, \hbar) = \exp\left(  \frac{1}{\hbar} S(x, \hbar) \right),
\end{equation}
where $S$ has an $\hbar$ expansion by itself
\begin{equation}
  S(x, \hbar) = \sum_{n=0}^\infty S_n(x) \hbar^n\,.
\end{equation}
We solve this equation order by order in $\hbar$.
This structure is very reminiscent of what we described in section \ref{sec:quantum_periods}.
The Schr\"odinger equation constructed there was solved by brane wave functions
and comparing this to \eqref{eqn:WKB_lead_ref} we see that to leading order we can 
identify 
\begin{equation}
  S_0(x) = - \int^x p(x') \rd x'
\end{equation}
so that the derivative of the leading order approximation of $S$ can be identified as being
the momentum
\begin{equation}
  S'_0(x) = - p(x)\,.
\end{equation}
Following this logic, we can use the derivative of $S$ to 
define a \emph{quantum differential} by setting
\begin{equation}
  \partial S = \partial_x S(x, \hbar) \rd x\, .
\end{equation}
But now we need to interpret the meaning of this quantum deformation and
in order to do that, we need to analyze the behavior of brane monodromies
on the Riemann surface.
We define the combination of $A$ and $B$ cylces of the Riemann surface
\begin{equation}
  \gamma_A = \sum_I \mathfrak{l}^I A_I, \quad \gamma_B = \sum_I \mathfrak{m}_I B^I
\end{equation}
around which we will move the branes.
These monodromies change the phase of the partition function as
\begin{equation}
   \CM_{\gamma_A}: Z_{\text{top}}(\vec{a}) \rightarrow \exp\left(\frac{1}{\epsilon_\alpha}  \sum_I \mathfrak{l}^I a_I  \right)Z_{\text{top}}(\vec{a})
\end{equation}
if we move the brane around the $A$-cycle while
it changes in the manner
\begin{equation}
 \CM_{\gamma_B}:\ Z_{\text{top}}(\vec{a}) \rightarrow Z_{\text{top}} (\vec{a} + \frac{g_s^2}{\epsilon_\alpha} \vec{\mathfrak{m}})
 = \exp\left( \frac{g_s^2}{\epsilon_\alpha} \sum_I \mathfrak{m}_I \frac{\partial}{\partial a_I} \right) Z_{\text{top}}(\vec{a})
\end{equation}
if we move the brane around the $B$-cycle.

The monodromy around $\gamma_B$ acts on $Z$ as a multiplication of
\begin{equation}
  \exp( \sum_I \frac{1}{\epsilon_\alpha} \mathfrak{m}_I a^I_D )
\end{equation}
so that a comparison yields
\begin{equation}
  a_D^I = g_s^2 \frac{\partial}{\partial a_I}\,.
\end{equation}
From observation made in \cite{Aganagic:2003qj}, we have 
\begin{equation}
	\Psi_2(x) = \langle e^{-\frac{1}{\hbar} \phi(x)} \rangle = e^{\frac{1}{\hbar} \int^x \partial S}
\end{equation}
and therefore
\begin{equation}
  Z_\text{top} (\vec{a}) \rightarrow e^{\frac{1}{\hbar} \oint_{\gamma_B} \partial S} Z_\text{top} (\vec{a})\,.
 \label{eqn:brane_mono}
\end{equation}
The partition function itself is given by
\begin{equation}
  Z_\text{top}  (\vec{a}; \epsilon_1, \epsilon_2 ) = 
  \exp \left( \sum_{g=0}^\infty g_s^{2g-2} \CF^{(g)}(\vec{a}; \hbar )  \right)
\end{equation}
which can be written as
\begin{equation}
  Z_{\text{top} }(\vec{a}; \epsilon_1 = 0, \epsilon_2 = \hbar) = \exp(   \CW(\vec{a}; \hbar))
\end{equation}
in the Nekrasov-Shatashvili limit.
We consider this as a deformation in $\hbar$ of the genus zero
amplitude of the unrefined topological string. As a result we can see now how the monodromy acts
on the partition function
\begin{equation}
  Z_\text{top}(\vec{a}) \rightarrow \exp \left( \sum_I \mathfrak{m}_I \partial_{a_I} \CW( \vec{a}; \hbar ) \right) Z_\text{top}(\vec{a}) \, .
\end{equation}
This has to be consistent with \eqref{eqn:brane_mono} and leads to the relations
\begin{equation}
  \oint_{B^I} \partial S = \partial_{a_I} \CW (\vec{a}; \hbar)
\end{equation}
where 
\begin{equation}
 \oint_{A_I} \partial S = a_I(\hbar)\,.
\end{equation}
These are $\hbar$ deformed quantum periods. This coincides with the special 
geometry relations presented in \ref{sec:special_geometry}.

Doing this suggests that we are able to extend special geometry to a quantum
deformed special geometry by lifting the classical periods to quantum periods 
by means of the quantum differential $\partial S$. We therefore define
\begin{equation}
  a_I (z_J; \hbar) = \oint_{A_I} \partial S \quad\text{and}\quad  a_D^I (z_J; \hbar) = \oint_{B_I} \partial S \quad I=1, \ldots, n \, ,
\end{equation}
which contain the classical periods as the leading order term of the semiclassical expansion.

The argument above leads us to conjecture that the relations between 
the quantum periods are just the common special geometry relations,
although with quantum deformed differential $\partial S$ and prepotential
$\CW(\hbar)$
\begin{equation}
  \frac{\partial \CW(\hbar)}{\partial a_I(z_J; \hbar)} = a_D^I(z_J; \hbar) \, .
  \label{eqn:SG_rel_pi}
\end{equation}

Using the WKB ansatz we plug \eqref{eqn:WKBansatz} into \eqref{eqn:HamiltonOperator}. 
This results in a sequence of $S'_n$, which are the corrections to the quantum periods
\begin{equation}
  a_I^{(n)} (z_J; \hbar) = \oint_{A_I} S'_{n}(x)\, \rd x \quad\text{and}\quad  {a_D^I}^{(n)} (z_J; \hbar) = \oint_{B_I} S'_{n}(x)\, \rd x, \quad I=1, \ldots, n \,.
\end{equation}

Another method to solve eq.~(\ref{eqn:HamiltonOperator}) is the use of so called \emph{difference equations} to solve for $\Psi$, which has been done in \cite{Aganagic:2011mi}.
This solves the problem perturbatively in the moduli $z_J$, while it is exact in $\hbar$. 
On the other hand the WKB ansatz is exact in the moduli $z_J$, while perturbative in $\hbar$.
Solving the Schr\"odinger equation via a difference equation is best shown 
by giving examples, which can be found in sections \ref{sec:conifold}, \ref{sec:localF0} and \ref{sec:localF1}.

At large radius the A-periods  can be expanded like
\begin{equation}
  a^{(n)} (z_J; \hbar) = \sum_{\vec{m} } 
  	\mathrm{Res}_{x = x_0 }  ( \partial_{\vec{m} }  S'_n( x ; z_J ) ) \frac{\vec{z}^{ \vec{m} } }{ \vec{m}! }
  	\label{eqn:largeRadiusExpansion}
\end{equation}
for $n>0$ and a suitably chosen point $x_0$. In the case $n = 0$ the leading order of the integrand has a branch cut so that we cannot just
take the residues.

After having explained how the WKB expansion is used, some comments about
the quantization of this system are in order.
The perturbative quantization condition for this problem is given by
(see \cite{Mironov:2009uv})
\begin{equation}
  \oint_B \partial S = 2 \pi \hbar \left( n + \frac{1}{2} \right)\quad n = 0, 1, 2, \cdots\,.
\end{equation}
However in \cite{Kallen:2013qla} it was shown that this is not a sufficient 
condition, because the B-periods have poles at infinitely many values
of the coupling constant. Hence this condition has to be extended to
a nonperturbative condition. The authors made the conjecture that
the nonperturbative part is actually controlled by the unrefined topological
string, somewhat dual to the observations made in \cite{Hatsuda:2013oxa}.

Another approach has been suggested in \cite{Krefl:2013bsa}, where
the condition
\begin{equation}
  \exp( \partial_{a_I} \CW( \vec{a}, \hbar  )) = 1
\end{equation}
was used as the starting point for defining a nonperturbative completion.
%
%
%
%
%
%
\subsection{Genus 1-curves}
\subsubsection{Elliptic curve mirrors and closed modular expressions}
\label{sec:ell_curves}
The next step would be to actually compute the genus zero amplitudes.
In order to do that a method has been developed in \cite{Huang:2011qx},
based on the work of \cite{KleinundFricke}\cite{Brandhuber:1996ng}.
The B-period is given in this formalism via the relation
\begin{equation}
  \frac{\partial}{\partial a} a_D(a,\vec{m} ) = - \frac{1}{2\pi \ri} \tau(t, \vec{m})
\end{equation}
and the prepotential $F^{(0,0)}$ can be calculated by making use of the
relation
\begin{equation}
  \frac{\partial^2}{\partial a^2} F^{(0,0)} (t,\vec{m} ) = - \frac{1}{2\pi \ri} \tau(a, \vec{m})
\end{equation}
between $F^{(0,0)}$ and the $\tau$-function of an elliptic curve.
This function is defined by 
\begin{equation}
  \tau = \frac{\int_b \omega}{\int_a \omega}
\end{equation}
where $a$ and $b$ are an integer basis of $H^1( \CC, \IZ )$ of the elliptic curve.

The elliptic curve needs to be given in Weierstrass form 
\begin{equation}
	y^2 = 4 x^3 - g_2(u,m) x - g_3(u, m)
\end{equation}
which is achieved by applying Nagell's algorithm.
Here $u$ is  the complex structure parameter of the curve and $m$ are isomonodromic deformations.

The local flat coordinate at a cusp point in the moduli space is the period 
over a vanishing cycle $\mu$. It can be obtained near such a 
point $u, \vec{m}$ by integrating
\begin{equation}
  \frac{\rd t}{\rd u} = \sqrt{ \frac{E_6(\tau) g_2(u, \vec{m}) }{ E_4(\tau) g_3(u, \vec{m}) } }\,.
\end{equation}
Here the functions $E_4$ and $E_6$ are the 
Eisenstein series.
Note that the $g_i$, while not invariants of the curve, can be rescaled by
\begin{equation}
 g_i \rightarrow \lambda^i(u, \vec{m}) g_i\ . 
\end{equation}
However the scaling function $\lambda(u, \vec{m})$ is very restricted by the 
requirement not to introduce new poles, zeros or branch cuts for the periods 
in the $u, \vec{m}$ parameter space. In practice the remaining freedom 
is easily fixed, by matching $\frac{\rd t}{\rd u}$ to the leading behaviour of the period 
integral at the cusp. E.g. near the large complex structure cusp, we match 
the leading behaviour    
\begin{equation}
  \frac{\rd t}{\rd u} = \frac{1}{2 \pi \ri} \int_\mu \frac{\rd x}{y} = \frac{1}{u} + \cdots \ . 
  \label{eqn:u_beh}
\end{equation}
and use the fact that the integration constant vanishes. This yields the 
period that is ususally called $a(u,\vec m)$ in Seiberg-Witten theory. 
Similarly at the conifold cusp, we can match similarly $t$ to the vanishing 
period $a_D(u,\vec m)$ at that cusp. 

We find the relation between $\tau$ and $t,\vec{m}$ by the fact that
the $j$ function has the universal behaviour
\begin{equation}
  j = 1728 \frac{E_4^3(\tau)}{E_4^3(\tau) - E_6^2(\tau)} = \frac{1}{q} + 744 + 196\ts{}884 q + 21\ts{}493\ts{}760 q^2 + \CO(q^3)
\end{equation}
where $q = \exp(2 \pi \ri \tau)$ which can then be inverted to obtain $\tau( j )$ .
The function $j$ on the other hand  can also be written in terms of  $t,\vec{m}$ 
\begin{equation}
  j = 1728 \frac{g_2^3(t, \vec{m})}{\Delta(t, \vec{m})}
\end{equation}
with $\Delta(t, \vec{m}) = g_2^3(t, \vec{m}) - 27 g_3^2(t, \vec{m}) $,
so that we can easily find an expression of $\tau$ in terms of $t, \vec{m}$.

With the formalism described above it is hence possible to write for all 
B-model curves and Seiberg Witten curves of genus one the classical vanishing 
period as well as the classical dual period (see~\cite{KleinundFricke,Brandhuber:1996ng} 
for more details regarding the latter period) at each cusp point. Alternatively 
one can write a differential operator, which is of third order in the derivatives w.r.t. 
$u$~\cite{Huang:2013yta} 
\begin{equation}
  {\cal D}^{(3)}(u, \vec m) \int_{a,b} \lambda =0\ .  
  \label{thirdorderoperator}
\end{equation}

%
%
%
%
%
\subsubsection{Special geometry}
In this article we are only concerned with Riemann surfaces of genus one.
As mentioned above this means effectively we only have two compact cycles. We will denote the periods around these
$a$ and $a_D$. The special geometry relation is given by
\begin{equation}
  a_D = \frac{\partial F}{\partial a} .
\end{equation}
At large radius we choose the periods in such a manner that we have a single logarithm 
in $u$ for the $a$-period, while we get squares of logarithms for the
$a_D$-period.
In this paper $\tilde{u}$ will correspond to the the compact
toric divisors inside the diagram.
Generally, we have to rescale it to find the moduli $u$ which
gives the leading $\log$-behaviour of the periods at large radius.
But we are considering local Calabi-Yau manifolds which generally have
additional non-normalizable parameters.
We will associate these parameters with the remaining noncompact toric divisors
and call them mass parameters, denoted by $m_i$
\begin{figure}[htp]
\begin{center}
\begin{minipage}{0.4 \textwidth}
\includegraphics[scale=0.2]{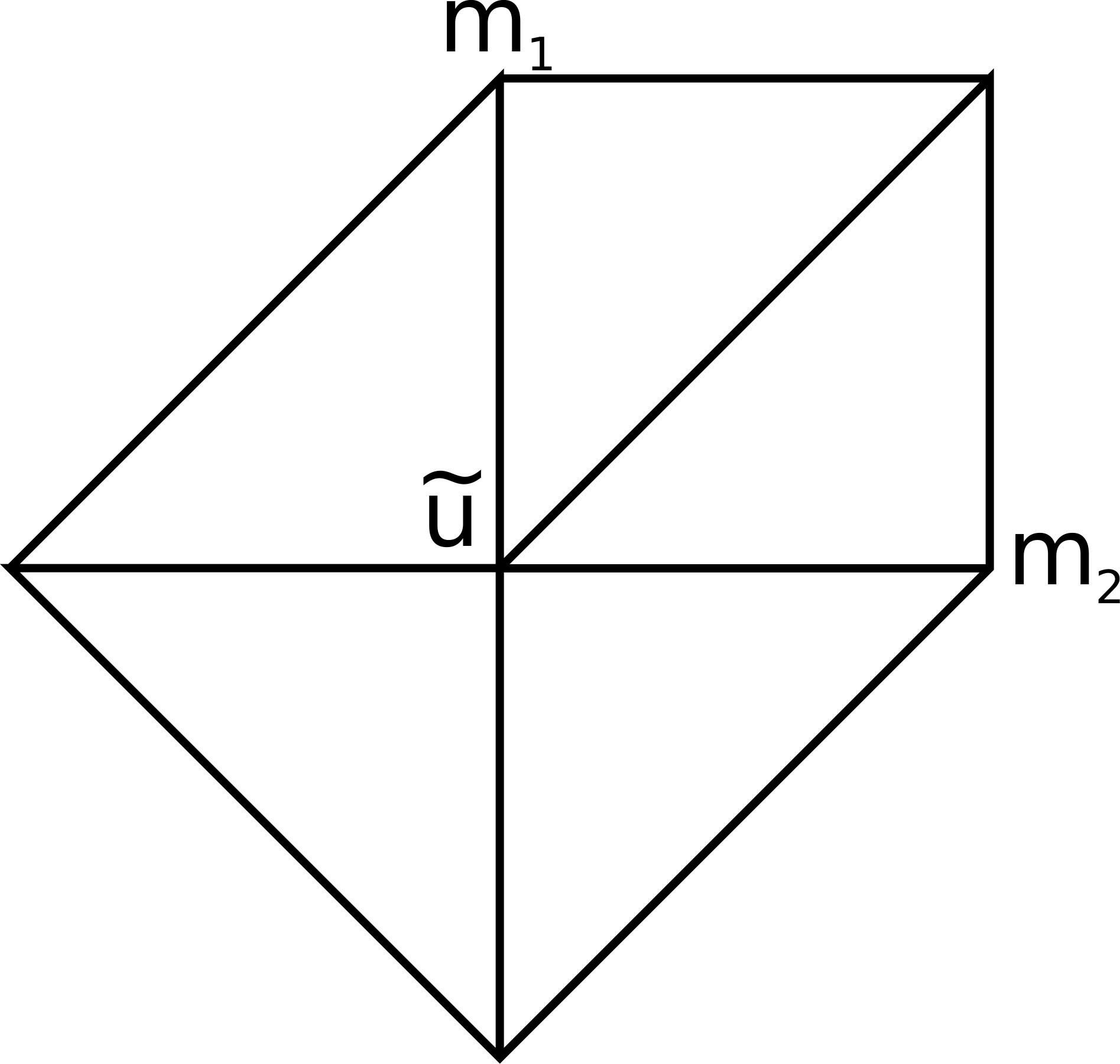}
\end{minipage}
\begin{minipage}{0.4 \textwidth}
$
 \label{datadp2_b} 
 \begin{array}{cc|crr|rrrl|c} 
    \multicolumn{5}{c}{\nu_i }    &l^{(1)} & l^{(2)} &l^{(3)}& &\\ 
    D_u    &&     1&     0&   0&           -1& -1& -1&  & \tilde{u}  \\ 
    D_1    &&     1&     1&   0&           -1&  1&  0& & m_2\\ 
    D_2    &&     1&     1&   1&            1& -1&  1& &\\ 
    D_3    &&     1&     0&   1&            0&  1& -1& & m_1\\ 
    D_4    &&     1&    -1&   0&            0&  0&  1& &\\ 
    D_5    &&     1&     0&  -1&            1&  0&  0& &
  \end{array} 
$
\end{minipage}
\end{center}
\caption{local $\mathcal{B}_2$. In the first column we denote the divisors
and in the fourth column the moduli and parameters associated with them.}
\label{fig:loc_B2_a1}
\end{figure}
Let us give an example. In figure \ref{fig:loc_B2_a1}, we have given
the data for local $\mathcal{B}_2$, which will be analyzed later on in 
\eqref{sec:localB2}. Here we have one normalizable moduli $\tilde{u}$ and
two mass parameters $m_1, m_2$. Looking at Batyrev's coordinates, we
find three coordinates
\begin{equation}
  z_1 = \frac{x_2 x_5}{x_0 x_1},\ z_2 = \frac{x_1 x_3}{x_0 x_2},\ z_3 = \frac{x_2 x_4}{x_0 x_3}
\end{equation}
where $x_0$ is associated with $D_u$ and therefore with $\tilde{u}$. Analogously for
the two mass parameters. Setting the remaining $x_i$ to one and defining $u = 1/\tilde{u}$,
we obtain the relation
\begin{equation}
  z_1 = \frac{u}{ m_2},\ z_2 = u m_1 m_2,\ z_3 = \frac{u}{ m_1}\,.
\end{equation}
The definition of $u$ follows from demanding the behaviour given in
\eqref{eqn:u_beh}. The operator $\Theta_u=u\partial_u$ can also be written in
terms of Batyrev coordinates which leads to
\begin{equation}
  \Theta_u =  \Theta_{z_1} + \Theta_{z_2} + \Theta_{z_3}\,.
\end{equation}
\subsubsection{Quantum Geometry}
In \cite{Vafa:2012fi} the connection between the dual toric diagrams
which show the base of an $T^2\times \IR$-fibration to $(p,q)$-branes
was interpreted. The result was that moving the external lines in $\IR^2$
requires an infinite amount of energy compared to the internal lines.
The degrees of freedom related to the external lines are the mass
parameters or non-normalizable moduli. Thus it makes
sense to consider them as being non-dynamical.
We assume that the quantum deformed periods remain non-dynamical,
meaning that they do not obtain quantum corrections and for genus one
only $a$ and $a_D$ will be quantum corrected.

As mentioned already, the quantum corrections to the periods
can be extracted from the meromorphic forms, derived by the WKB ansatz
which we use to solve the Schr\"odinger equation.
For the A-periods, this reduces to residues, except for the logarithmic
part of the classical contribution. Often it is possible to match
the contributions from the residues to the different A-periods,
but in some examples even this is not easily possible. For the B-periods it is even harder,
because we generally have to find different parameterizations, giving
different contributions, which have to be summed up in order to find
the full result.
The local $\IP^2$ like it was solved in \cite{Aganagic:2011mi}, is a 
good example for this,
as well as the local $\IF^1$, see section \ref{sec:F1_diff_eq}.
Actually, this problem even appears for the local $\IF_0$ 
(see section \ref{sec:localF0}, but because it is very symmetric we
do not actually have to do any additional computations in order to
solve this problem).

We want to avoid this complications, therefore we use a different approach.
It is possible to derive differential-operators that give quantum corrections 
by acting on the classical periods. It turns out, that these operators
are only of second order. 

Having found these operators, the strategy is to apply them to the solution
of the Picard-Fuchs system and build
up the quantum corrections. The idea is that the operator is exact
under the period integral, so that we can use partial integration to 
derive it.

The quantum periods $a(u,\vec{m}; \hbar)$ will be build up from the classical one $a(u,\vec{m})$ in the manner
\begin{equation}
  a(u,\vec{m};\hbar) =[1 + \sum_{i=1}^\infty \hbar^{2 i} \CDD_{2 i}]a(u,\vec{m})=: {\cal D}^{(2)}(u, \vec{m},\hbar)    a(u,\vec{m})\, .
\end{equation}
The individual $\CDD_{i}$ are second order differential operators in $u$
given by
\begin{equation}
 \CDD_{i} = a_i ( u, \vec{m} ) \Theta_u + b_i ( u, \vec{m} ) \Theta_u^2
\end{equation}
where $\Theta_u = u \partial_u$ and  $a_i ( u, \vec{m} )$ and $b_i ( u, \vec{m} )$ are rational functions in their arguments. 
We do not have proven that this is always true, but for the examples we considered 
it has always been a viable ansatz. We derive such operators by taking the full WKB function under
an integral with a closed contour and then applying partial integration.

The same holds for the dual period
\begin{equation}
  a_D(u,\vec{m};h) ={\cal D}^{(2)}(u, \vec{m},h)    a_D(u,\vec{m})\ .
\end{equation}

This approach has been introduced in \cite{Mironov:2009uv}
and used in \cite{Aganagic:2011mi} for the geometry
corresponding to a matrix model with a cubic potential.
It also has been applied to the local $\IF_0$ and local $\IP^2$ in \cite{Huang:2012kn}.
We are going to apply in even more examples while assuming that the operator
is, at least at order $\hbar^2$, always of order two. It would be very
interesting to provide a proof for this conjecture.
%
%
%
%
%
\section{Examples}
\label{sec:computations}
\subsection{The resolved Conifold} \label{sec:conifold}
Let us start with a simple example, namely the resolved conifold.
Its charge-vector is given by
\begin{equation}
  l = (-1, -1, 1, 1) \, .
\end{equation}
Using the given charge vector we find for the Batyrev coordinate
\begin{equation}
z=\frac{x_3 x_4}{x_1 x_2}
\end{equation}
which leads to the mirror curve 
\begin{equation}
  u v = 1 + e^x + e^p + z e^x e^{-p} \, .
\end{equation}
Due to the adjacency of $x$ and $p$ in the last term of the sum
we have quantum corrections in the Hamiltonian, 
due to the normal ordering ambiguities. The quantum Hamiltonian is
\begin{equation}
  H = 1 + e^x + e^p + z e^{- \hbar/2} e^x e^{-p} \,.
\end{equation}
This Hamiltonian leads to the difference equation
\begin{equation}
  V(X) = -1 - X - z \frac{ X e^{- \hbar/2} }{ V(x - \hbar) }
\end{equation}
where $X= e^x$ and $V(x) = \psi(x+\hbar)/\psi(x)$.
The $A$-period does not obtain any quantum corrections and is therefore given by
\begin{equation}
  a = \log( z ) \, .
\end{equation}
After defining $Q=e^a$ we can invert this and find for the mirrormap $z = Q$.
The $B$-period up to the fourth order in $Q$ is
\begin{equation}
 \tilde{a}_D = \frac{q^{1/2} \log q}{q - 1} Q + \frac{1}{2} \frac{q \log q}{q^2 - 1} Q^2 + \frac{1}{3} \frac{q^{3/2} \log q}{q^3 - 1} Q^3  + \frac{1}{4} \frac{q^{2} \log q}{q^4 - 1} Q^4 + \CO(Q^5)
 \label{eqn:rescon_atil}
\end{equation}
where $q=e^{\hbar}$.
The structure is very suggestive and leads us to assume the full form to be
\begin{equation}
  a_D = \log q \sum_{i=1}^\infty  \frac{1}{i} \frac{Q^i}{q^{i/2} - q^{-i/2}}  .
\end{equation}

The resolved conifold does not have a compact B-cycle and therefore only
has a mass-parameter. But according to \cite{Forbes:2005xt} the 
double-logarithmic solution can be generated by the Frobenius
method. The fundamental period for the resolved conifold is
\begin{equation}
 \varpi(z; \rho) = \sum_{n=0}^\infty \frac{z^{n+\rho}}{\Gamma (1-n-\rho)^2 \Gamma (1+n +\rho )^2}
\end{equation}
and 
\begin{equation}
  \partial_\rho^2 \varpi(z; \rho) \Big|_{\rho = 0}= \log^2(z)+2 z+\frac{z^2}{2}+\frac{2 z^3}{9}+\frac{z^4}{8} + \CO(z^5)
\end{equation}
generates the B-period. The non-logarithmic part of this is indeed given by the semiclassical limit of \eqref{eqn:rescon_atil}.
Therefore, we define
\begin{equation}
  a_D = \frac{1}{2} \log^2(z) + \tilde{a}_D
\end{equation}
and use this formally as our dual period.
Integrating the special geometry relations gives us the free energy in the Nekrasov-Shatashvili limit 
\begin{equation}
 \CW = \hbar \sum_{i=1}^\infty  \frac{1}{i^2} \frac{Q^i}{q^{i/2} - q^{-i/2}} \,.
\end{equation}
The full free energy can by computed via the refined topological vertex and is
\begin{equation}
  F_{\text{RTV}} = - \sum_{i=1}^\infty \frac{1}{i}\frac{Q^i}{(q^\frac{i}{2} - q^{-\frac{i}{2}})(t^\frac{i}{2} - t^{-\frac{i}{2}})}
\end{equation}
with 
\begin{equation}
  q = e^{\epsilon_1}\quad \text{and}\quad t = e^{\epsilon_2} .
\end{equation}
The Nekrasov-Shatashvili limit is defined in \eqref{eqn:NS_limit} and
plugging in the free energy of the refined conifold into this, we find
\begin{equation}
  \CW_{\text{RTV}} =-  \epsilon_2 \sum_{i=1}^\infty \frac{1}{i}\frac{Q^i}{t^\frac{i}{2} - t^{-\frac{i}{2}}}\left( \lim_{\epsilon_1 \rightarrow 0} \frac{\epsilon_1}{q^\frac{i}{2} - q^{-\frac{i}{2}}} \right) = 
  -  \epsilon_2 \sum_{i=1}^\infty \frac{1}{i^2}\frac{Q^i}{t^\frac{i}{2} - t^{-\frac{i}{2}}}
\end{equation}
which is exactly the result we found using the quantum periods.
This also fits nicely in the expansion presented in \eqref{eqn:Wexp} with the only
nonvanishing \emph{instanton number} being $\hat{n}_0 = 1$.
%
%
%
%
%
\subsection{\texorpdfstring{local $\IF_0$}{local F0} }
\label{sec:localF0}
\begin{figure}[h]
\begin{center}
\includegraphics[scale=0.5]{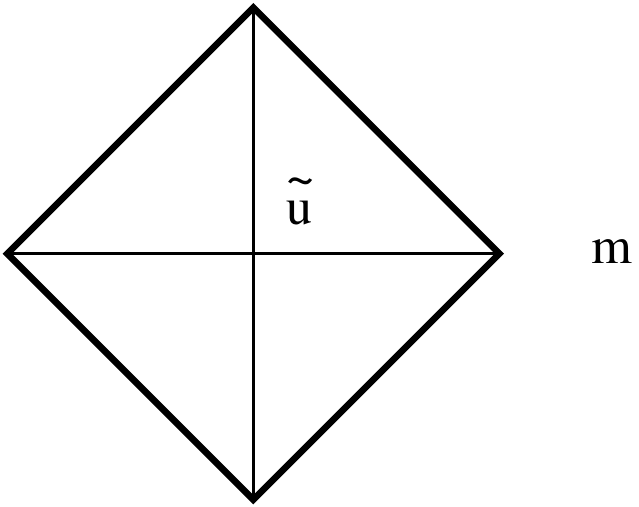}
\caption{\label{fig:poly2} Polyhedron 2 depicting the toric geometry $\mathds{F}_0$.}
\end{center}
\end{figure}

We begin with presenting the Mori cone for the toric geometry depicted in fig.~(\ref{fig:poly2})
\begin{equation} 
 \label{dataf0} 
 \begin{array}{cc|rrr|rrl|} 
    \multicolumn{5}{c}{\nu_i }    &l^{(1)}& l^{(2)}&\\ 
    D_u    &&     1&     0&   0&         -2&  -2&       \\ 
    D_1    &&     1&     1&   0&         1&   0&        \\ 
    D_2    &&     1&     0&   1&         0&   1&        \\ 
    D_3    &&     1&    -1&   0&         1&   0&        \\ 
    D_4    &&     1&     0&   -1&        0&   1&       \\  
  \end{array} \ . 
\end{equation} 
From the toric data we find the complex structure moduli at the large radius point
\begin{equation}\label{eq:f0mpara}
z_1=\frac{m}{\tilde{u}^2}\, ,\quad z_2 = \frac{1}{\tilde{u}^2} \, .
\end{equation}
After setting $u=\frac{1}{\tilde{u}^2}$ the mirror curve in these coordinates is given by
\begin{equation} 
  -1 + e^x + e^p + m u\, e^{-x} + u\, e^{-p} = 0 \, .
  \label{eqn:P1tP1curve}
\end{equation}
Hence the Schr\"odinger equation for the brane wave function corresponding to this reads
\begin{equation}
    (-1 + e^x + mu\, e^{-x} ) \Psi(x) + \Psi(x + \hbar) + u\, \Psi(x-\hbar) = 0 \, .
\end{equation}
The coefficients of the classical Weierstrass normal form are
\begin{subequations}
\begin{align}
g_2(u, m) &= 27 u^4 (1 - 8 u - 8 m u + 16 u^2 - 16 m u^2 + 16 m^2 u^2) \, ,\\
g_3(u, m) &= 27 u^6 (1 - 12 u - 12 m u + 48 u^2 + 24 m u^2 + 48 m^2 u^2 \nonumber \\
& \phantom{={}} - 64 u^3 + 96 m u^3 + 96 m^2 u^3 - 64 m^3 u^3) \, .
\end{align}
\end{subequations}
\subsubsection{Difference equation}
Defining the function
\begin{equation} \label{eqn:defV}
 V(x) = \frac{\Psi(x + \hbar)}{\Psi(x)},
\end{equation}
we obtain the difference equation
\begin{equation}
 	V(x) = 1 - e^x + mu\, e^{-x} + \frac{u}{V(x - \hbar)},
\end{equation}
which can be expanded around $u = 0$.
Doing this leads to a power series for $V(x)$
\begin{equation}
  V(x) = 1 - X + m u\, e^{-x} + \frac{u}{1 - q^{-1} e^x} + \CO(u^2),
\end{equation}
where we defined $q = e^\hbar$.

We see in \eqref{eqn:defV} that $V$ includes the wavefunction $\psi$, which includes the quantum 
differential we seek. 
This has been already used in section \ref{sec:conifold}, when we solved
the resolved conifold, but let us see how we can actually extract it from 
this expression by integrating over $\log{V}$
\begin{equation}
  \int \log(V(x)) = \int S'(x) \rd x + \sum_{n=2}^\infty \frac{\hbar^n}{n!} \int S^{(n)}(x) \rd x 
\end{equation}
where we used $\Psi(x)=e^{\frac{1}{\hbar}\sum_{i=0}^\infty S_i(x) \hbar^i}$.
Integrating around a closed contour the last part vanishes because we have
for $ n \geq 2$
\begin{equation}
   \oint \rd x S^{(n)}(x) = \left[ S^{(n-1)} \right] \,,
\end{equation} 
so that indeed
\begin{equation}
 \oint \partial S = \oint \log(V(X))
\end{equation}
and we can use this quantity to define the quantum differential for closed
contours.

Computing the contour integral around the $A_I$ cycles leads to

\begin{align}
	\frac{1}{2 \pi \ri}\oint_{A_I} \log(V(x; z_I)) \rd x 
	& = \frac{1}{2 \pi \ri}\oint_{A_I} \log(V(x; 0)) \rd x
	+ \sum_{ \vec{n} } \frac{1}{\vec{n}!}  \mathrm{Res}_{X = X_0} \frac{1}{X} \partial_{ {\vec{z}}^{\vec{n}} } \log(V(X; \vec{z} ) ) \\
	& \propto \log( z_I ) + \tilde{a}   \label{eqn:aper_P1tP1} 
\end{align}
where we defined $X = e^x$ and $X_0$ is a appropriately chosen pole.

In this case the A-Period is is given by
\begin{equation}
 a = \log u + 2 (m+1) u +  \left(3 m^2+2 m \left(q+\frac{1}{q}+4\right)+3\right)u^2 + \CO(u^3)
\end{equation}

The B-periods are more complicated to obtain. Due to the symmetry of this case, we find the contributions
to the B-period by taking an integral and symmetrizing with respect to
$u \leftrightarrow u m$.
This is due to the symmetry of local $\IF_0$ and as a result we only have 
to do the integration once in order to obtain the final result.

We regularize the integrals using the boundaries $\delta$ and $\Lambda$. Using these we
extract the finite (and real) part of the integral. 

The $B$-period of local $\IF_0$ is given by
\begin{align}
 a_D & = -\frac{1}{2} \log (u) \log (m u) -\frac{1}{2} \log \left(m u^2\right) a
 + \tilde{a}_D
\end{align}
where the non-logarithmic part of the $B$-periods is given by
\begin{align}
 \tilde{a}_\text{int} &  = \int_\delta^\Lambda \log(V(x)) \rd x \\
 & = 4 \left( \frac{q+1}{q-1}\log{q}\right) z_2 + 4 z_1^2 + \left(4 + 2 \frac{\left(5 q^2+8 q+5\right) \log{q}}{q^2-1} \right)z_2^2 \nonumber \\
 & \phantom{={}} + \left(8 + 4 \frac{ (q+1)^3 }{(q-1) q}\log{q}) \right)z_1 z_2 + \CO(z_i^3)\;.
\end{align}
In this expression we had to symmetrize with respect to $\hbar \rightarrow -\hbar$ to get rid of the odd sector in $\hbar$. That is expected because we only integrated over
a small portion of the surface and are going to piece together the full period
by symmetry considerations.
Symmetrizing with respect to the variables $u$ and $u m$ finally yields
\begin{align}
 \tilde{a}_D & = -\frac{(m+1) (q+1)  \log (q)}{q-1} u  
 -2 (1 + m)^2 u^2 + \nonumber \\
 & \phantom{={}} -\frac{\left(m^2 q \left(5 q^2+8 q+5\right)+4 m (q+1)^4+q \left(5 q^2+8 q+5\right)\right) \log (q)}{2 q \left(q^2-1\right)} u^2 
 + \CO(u^3) \, .
\end{align}

Due to the symmetry of local $\IF_0$ we only had to compute one integral. 
But generally we would have to find the right parameterizations and piece together 
the results to yield the full periods. This will also become an issue for the 
A-periods in more complicated cases.
We define the single valued quantity 
\begin{equation}
  Q_t = \exp( a )\,.
\end{equation}
The special geometry relations in this variable are
\begin{equation}
   Q_t \partial_{Q_t} \CW(Q_1, Q_2; q) = \tilde{a}_D (Q_t, Q_m; q)
\end{equation}
which yields
\begin{equation}
  \CW (Q_1, Q_2; q) = \frac{1 + q}{1 - q}(Q_1 + Q_2) + \frac{1}{4}\frac{1 + q^2}{1 - q^2}(Q_1^2 + Q_2^2)
  + \frac{(1 + q)(1 + q^2)}{q(1 - q)} Q_1 Q_2 + \CO(Q_i^3)\ ,
\end{equation}
if we drop the classical terms.

One advantage of this method is that it is exact in $\hbar$ from the beginning.
In the following cases it will be very hard though to extract the correct contributions
to the periods from the quantum differential form.  The information we find
due to this method can be obtained in any case, because at the large radius point
the BPS numbers have the property
\begin{equation}
 N^\beta_{j_L, j_R} = 0\, \text{for}\, \beta > \beta^\text{max}(j_L, j_R) 
\end{equation}
for finite $\beta^\text{max}(j_L, j_R) $.
As a result we can reconstruct the data, found in this section by a computation
perturbative in $\hbar$.

If we want to compute amplitudes at different point in moduli space we also
have the problem that it is quite hard to compute the amplitudes exactly in $\hbar$,
because we cannot properly attribute the contours. Hence, for the next
section, we will use another approach, namely the differential operators,
which map the classical periods to the higher order corrections.
\subsubsection{Operator approach}
From the Schr\"odinger equation given above we find for the zeroth order WKB function
\begin{equation}
S_0^\prime (x) = \log \left(-\frac{e^{-x}}{2}\left(-e^x+e^{2 x}+m u+\sqrt{(-e^x+e^{2 x}+m u)^2-4 e^{2 x} u}\right) \right) \, .
\end{equation}
The operator mapping the zeroth to the second order periods is given by
\begin{equation}
 \CDD_2 = \frac{1}{6} (-u - m u) \Theta_u + \frac{1}{12} (1 - 4 u - 4 m u)\Theta_u^2 \, ,
\end{equation}
with $\theta_u=u\partial _u=z_1\partial _1+z_2\partial _2$ in terms of the Batyrev coordinates. Higher order operators are given in appendix \ref{app:localF0}. The form of $\theta_u$ already leads to the conclusion that the $m$ parameter defined in eq.~(\ref{eq:f0mpara}) does not get any $\hbar$ corrections like it was expected for trivial parameters. 
Computing the classical A-period as described in section \ref{sec:ell_curves} and using the operators to calculate quantum corrections gives the expression
\begin{subequations}
\begin{align}
 a &= \log(u) + 2 (1 + m) u + 3 (1 + 4 m + m^2) u^2  \nonumber \\
 &\phantom{=}{} + \frac{20}{3} (1 + 9 m + 9 m^2 + m^3)u^3 + \left (2 m u^2 + 20 m (1 + m) u^3 \right) \hbar^2 + \CO(\hbar^3, u^4)
\end{align}
and the B-period is given by
\begin{align}
a_D & = - 2 \log(u)^2 - 2 \log(m) \log(u)  \nonumber \\
&\phantom{={}} - 2 \log(m u^2) \left ( 2 u (1+m) + 3 u^2 (1+4m+m^2) + 2 m u^2 \hbar^2  \right ) - 8 u(1+m) \nonumber \\
&\phantom{={}} - 2 u^2 (13 + 40m + 13m^2) - \frac{1}{3} \hbar^2 (1 + 2 u (1+m) + 2u^2 (3 +32m+3m^2) ) + \CO(\hbar^3, u^3)\, .
\end{align}
\end{subequations}
Inverting the exponentiated A-period we find for the mirror map
\begin{align}
u(Q_u) &= Q_u - 2 (1 + m) Q_u^2 + 3 Q_u^3 (1+m^2) -4Q_u^4(1+m+m^2+m^3) \nonumber \\
&\phantom{= Q_u - 2 (1 + m) Q_u^2 + 3 Q_u^3 (1+m^2)} {} + (-2 m Q_u^3 - 4 m (1 + m) Q_u^4) \hbar^2 + \CO(\hbar^3, Q_u^5)\, .
\end{align}
To integrate the special geometry relation we need the following relations
\begin{equation}
Q_u=Q_2 ,\; m=\frac{Q_1}{Q_2} \, ,
\end{equation}
which can be checked by calculating the periods as solutions of the Picard-Fuchs equations. Then we find the instanton numbers given in the tables \ref{tab:instf0genus0}, \ref{tab:instf0genus1}, \ref{tab:instf0genus2} and \ref{tab:instf0genus3}.

\begin{table}[h!]
\begin{center}
\begin{tabular}{c|ccccccc} 
 & $d_1$ & 0 & 1 & 2 & 3 & 4 & 5  \\
\hline
 $d_2$ & & & & & & &  \\
0 &  &   & $-2$ &   &   &   &  \\
1 &  & $-2$ & $-4$ & $-6$ & $-8$ & $-10$ & $-12$\\
2 &  &   & $-6$ & $-32$ & $-110$ & $-288$ & $-644$\\
3 &  &   & $-8$ & $-110$ & $-756$ & $-3556$ & $-13072$\\
4 &  &   & $-10$ & $-288$ & $-3556$ & $-27264$ & $-153324$\\
5 &  &   & $-12$ & $-644$ & $-13072$ & $-153324$ & $-1252040$\\
\end{tabular}
\caption{\label{tab:instf0genus0}The instanton numbers for local $\mathds{F}_0$ at order $\hbar^0$.}
\end{center}
\end{table}

\begin{table}[h!]
\begin{center}
\begin{tabular}{c|ccccccc} 
 & $d_1$ & 0 & 1 & 2 & 3 & 4 & 5  \\
\hline
 $d_2$ & & & & & & &  \\
0 &  &   & $-1$ &   &   &   &  \\
1 &  & $-1$ & $-10$ & $-35$ & $-84$ & $-165$ & $-286$\\
2 &  &   & $-35$ & $-368$ & $-2055$ & $-7920$ & $-24402$\\
3 &  &   & $-84$ & $-2055$ & $-21570$ & $-142674$ & $-699048$\\
4 &  &   & $-165$ & $-7920$ & $-142674$ & $-1488064$ & $-10871718$\\
5 &  &   & $-286$ & $-24402$ & $-699048$ & $-10871718$ & $-113029140$\\
\end{tabular}
\caption{\label{tab:instf0genus1}The instanton numbers for local $\mathds{F}_0$ at order $\hbar^2$.}
\end{center}
\end{table}

\begin{table}[h!]
\begin{center}
\begin{tabular}{c|ccccccc} 
 & $d_1$ & 0 & 1 & 2 & 3 & 4 & 5  \\
\hline
 $d_2$ & & & & & & &  \\
0 &  &   &   &   &   &   &  \\
1 &  &   & $-6$ & $-56$ & $-252$ & $-792$ & $-2002$\\
2 &  &   & $-56$ & $-1352$ & $-12892$ & $-75016$ & $-322924$\\
3 &  &   & $-252$ & $-12892$ & $-219158$ & $-2099720$ & $-13953112$\\
4 &  &   & $-792$ & $-75016$ & $-2099720$ & $-30787744$ & $-298075620$\\
5 &  &   & $-2002$ & $-322924$ & $-13953112$ & $-298075620$ & $-4032155908$\\
\end{tabular}
\caption{\label{tab:instf0genus2}The instanton numbers for local $\mathds{F}_0$ at order $\hbar^4$.}
\end{center}
\end{table}

\begin{table}[h!]
\begin{center}
\begin{tabular}{c|ccccccc} 
 & $d_1$ & 0 & 1 & 2 & 3 & 4 & 5  \\
\hline
 $d_2$ & & & & & & &  \\
0 &  &   &   &   &   &   &  \\
1 &  &   & $-1$ & $-36$ & $-330$ & $-1716$ & $-6435$\\
2 &  &   & $-36$ & $-2412$ & $-41594$ & $-375052$ & $-2288546$\\
3 &  &   & $-330$ & $-41594$ & $-1209049$ & $-17227788$ & $-157648036$\\
4 &  &   & $-1716$ & $-375052$ & $-17227788$ & $-365040880$ & $-4760491974$\\
5 &  &   & $-6435$ & $-2288546$ & $-157648036$ & $-4760491974$ & $-85253551830$\\
\end{tabular}
\caption{\label{tab:instf0genus3}The instanton numbers for local $\mathds{F}_0$ at order $\hbar^6$.}
\end{center}
\end{table}
\subsubsection{Orbifold point}
In this section we solve the problem at the orbifold point of the moduli space $\CM$.
This is very useful because it would be the point where we could compare our results
to a matrix model description of the refined topological string.

The coordinates around we want to expand at the orbifold point are given by \cite{Haghighat:2008gw}
\begin{equation}
  x_1 = 1 - \frac{z_1}{z_2},\; x_2 = \frac{1}{\sqrt{z_2}\left(1 - \frac{z_1}{z_2} \right)}.
\label{eqn:orbifold_coord}
\end{equation}
Due to the symmetry of local $\IP^1 \times \IP^1$ obtaining the quantum
periods has not been very hard in the large radius case. 
Now we want to find the quantum periods, expanded in the coordinates at the orbifold point.
The problem we are facing in this case is, that we do not know how to extract the relevant parts
from the integrals over the quantum differential. Hence we have to pursue a different
path in order to compute the quantum periods. In \cite{Huang:2012kn} certain operators
have been derived which allow us to obtain the higher order corrections in $\hbar$ via
second order differential operators, acting on the classical periods. We list
the operators for the second and fourth order here
\begin{subequations}
\begin{align}
t_2 &= -\frac{z_1+z_2}{6} \Theta_u t + \frac{1-4z_1-4z_2}{12} \Theta_u^2 t \label{eqn:p1tp1_t2} \\
t_4 &= \frac{1}{360\Delta^2} \{ 2[  z_1^2 (1 - 4 z_1)^3 + z_2^2 (1 - 4 z_2)^3+4z_1z_2(8 - 37 z_1  
			- 37 z_2 - 328 z_1^2 + 1528 z_1 z_2  \nonumber \\ 
    & \phantom{=} - 328 z_2^2  + 1392 z_1^3 -  1376 z_1^2 z_2  - 1376 z_1 z_2^2 + 1392 z_2^3)]\Theta_u t
       +[  -z_1 (1 - 4 z_1)^4 - z_2 (1 - 4 z_2)^4 \nonumber \\ 
    & \phantom{=} +4z_1z_2(69 - 192 z_1 - 192 z_2 - 1712 z_1^2 + 6880 z_1 z_2 - 1712 z_2^2 + 5568 z_1^3  
    	 -  5504 z_1^2 z_2 \nonumber \\ 
    & \phantom{=} - 5504 z_1 z_2^2 + 5568 z_2^3 )]\Theta_u^2 t \}  \label{eqn:p1tp1_t4}.
\end{align}
\end{subequations}
In the coordinates $z_1, z_2$ the logarithmic derivative $\Theta_u$ is
given by
\begin{equation}
  \Theta_u = u \partial_u = z_1 \partial_{z_1} + z_2 \partial_{z_2} \, .
\end{equation}
We can transform this operator to the orbifold coordinates
and act with it on the classical period. The expansion of classical periods
in the orbifold coordinates can be computed via solving the Picard-Fuchs
system. This has been done in \cite{Aganagic:2002wv} already so
that we can build on the solutions already at hand.

Here we present the periods to zeroth order 
\begin{subequations}
\label{eqn:periodslocF0OP}
\begin{align}
 \omega_0 & = 1 , \\
 s_1^{(0)} & = - \log(1 - x_1) = t_1 - t_2 \label{eqn:p1tp1_s1} \\
 s_2^{(0)} & = x_1 x_2 + \frac{1}{4} x_1^2 x_2 + \frac{9}{64} x_1^3 x_2 + \CO(x_i^4) \label{eqn:p1tp1_s2} \\
 F^{(0)}_{s_2} & = \log(x_1) s_2+  x_1 x_2 + \frac{3}{4} x_1^2 x_1 + \frac{15}{32} x_1^3 x_2 - \frac{1}{6} x_1 x_2^3 + \CO(x_i^4) \, . \label{eqn:p1tp1_F}
\end{align}
\end{subequations}
Using \eqref{eqn:p1tp1_t2} and \eqref{eqn:p1tp1_t4} we are able to compute the periods at order two and four, respectively.
As explained, we take the operators \eqref{eqn:p1tp1_t2} and \eqref{eqn:p1tp1_t4},
corresponding to order two and four, respectively, change the coordinates via \eqref{eqn:orbifold_coord} 
and apply them to \eqref{eqn:periodslocF0OP} to find
\begin{align*}
  s_2^{(2)} & = \frac{1}{64} x_1 x_2 + \frac{1}{256} x_1^2 x_2 + \frac{15}{8192} x_1^3 x_2 + \frac{35}{32768} x_1^4 x_2 + \CO(x_i^5)
\end{align*}
and
\begin{align}
  F^{(2)}_{s_2} & = -\frac{1}{8} x_2 + \frac{1}{6} \frac{x_2}{x_1} + \frac{3}{128} x_1 x_2 + 
  \frac{17}{1536} x_1^2 x_2 + \frac{189}{32768} x_1^3 x_2 + \frac{1387}{393216} x_1^4 x_2 \nonumber \\
  	&\phantom{={}}  - \frac{5}{24} x_2^3 + \frac{1}{6} \frac{x_2^3}{x_1} + 
  \frac{5}{128} x_1 x_2^3 + \frac{5}{1536} x_1^2 x_2^3 - \frac{7}{24} x_2^5 + \frac{1}{6} \frac{x_2^5}{x_1} \nonumber \\
  & \phantom{={}} + s_2^{(2)} \log (x_1)  + \CO(x_i^5)
\end{align}
while $s_1^{(0)}$ will not get any quantum corrections at all.

The result we find be using special geometry is
\begin{subequations}
\begin{align}
F^{(1,0)} & = \frac{1}{24}\left( \log(S_1) + \log(S_2) \right) + \frac{1}{576}( S_1^2+30 S_2 S_1+S_2^2 ) \nonumber \\
& \phantom{ = {} }-\frac{1}{138\ts{}240}( 2 S_1^4-255 S_2 S_1^3+1530 S_2^2 S_1^2 + S_1 \leftrightarrow S_2 ) \nonumber \\ 
& \phantom{ = {} } +\frac{1}{34\ts{}836\ts{}480} ( 8 S_1^6+ 945 S_2 S_1^5 - 43\ts{}470 S_2^2 S_1^4+150\ts{}570 S_2^3 S_1^3 + S_1  \leftrightarrow S_2 )+ \CO(S^7)
\end{align}
\begin{align}
F^{(2,0)} & = -\frac{7}{5760}\left( \frac{1}{S_1^2} + \frac{1}{S_2^2} \right) + \frac{1}{2\ts{}438\ts{}553\ts{}600}( 155 S_1^2-16\ts{}988\ts{}774 S_2 S_1+155 S_2^2 ) \nonumber \\
& \phantom{ = {} }-\frac{1}{5\ts{}573\ts{}836\ts{}800}( 31 S_1^4+13\ts{}093\ts{}484 S_2 S_1^3-27\ts{}178\ts{}854 S_2^2 S_1^2 + S_1 \leftrightarrow S_2 ) \nonumber \\ 
& \phantom{ = {} } -\frac{1}{14\ts{}714\ts{}929\ts{}152\ts{}000} ( 4960 S_1^6+\ts{}3842\ts{}949\ts{}687 S_2 S_1^5-36\ts{}703\ts{}156\ts{}395 S_2^2 S_1^4 \nonumber \\
& \phantom{ = {}  -\frac{1}{14\ts{}714\ts{}929\ts{}152\ts{}000} (}
{}+82\ts{}152\ts{}486\ts{}440 S_2^3 S_1^3 + S_1  \leftrightarrow S_2 ) + \CO(S^7)
\end{align}
\begin{align}
F^{(3,0)} & = -\frac{31}{161\ts{}280} \left( \frac{1}{S_1^4} + \frac{1}{S_2^4} \right)  - \frac{1}{1\ts{}560\ts{}674\ts{}304\ts{}000}( 2667 S_1^2-3\ts{}669\ts{}924\ts{}266 S_2 S_1+2667 S_2^2 ) \nonumber \\
& \phantom{ = {} }+\frac{1}{1\ts{}961\ts{}990\ts{}553\ts{}600}( 508 S_1^4+4\ts{}960\ts{}681\ts{}415 S_2 S_1^3-6\ts{}516\ts{}516\ts{}390 S_2^2 S_1^2  + S_1 \leftrightarrow S_2 ) \nonumber \\ 
& \phantom{ = {} } -\frac{1}{80\ts{}343\ts{}513\ts{}169\ts{}920\ts{}000} (1\ts{}930\ts{}654 S_1^6-6\ts{}435\ts{}720\ts{}4136\ts{}601 S_2 S_1^5+346\ts{}657\ts{}135\ts{}824\ts{}060 S_2^2 S_1^4 \nonumber \\
& \phantom{ = -\frac{1}{80\ts{}343\ts{}513\ts{}169\ts{}920\ts{}000}  ) } {} -727\ts{}232\ts{}136\ts{}215\ts{}170 S_2^3 S_1^3  + S_1  \leftrightarrow S_2 ) + \CO(S^7)
\end{align}
\end{subequations}
where the constants of integration have been fixed in a manner as to
give the resolved Conifold if we send $S_1 \rightarrow 0$ or $S_2 \rightarrow 0$.
The data before fixing the constant of integration can be found in appendix \ref{app:F0_orbifold_point}.

The relation between the periods in this case and the 't Hooft parameters of the corresponding
matrix model is, at least in the unrefined case, given by \cite{Aganagic:2002wv}
\begin{equation}
  S_1 = \frac{1}{4} ( s_1 + s_2 ),\quad S_2 = \frac{1}{4}(s_1 - s_2) \, .
\end{equation}
Hence the periods in terms of the 't Hooft parameters are given by
\begin{equation}
  s_1 = 2(S_1 + S_2) ,\quad s_2 = 2(S_1 - S_2) \,.
  \label{eqn:stoS}
\end{equation}

The case of $S_1 = 0$ or $S_2 = 0$ specializes to the resolved conifold at the orbifold point
which can be easily computed. Using this result, we are able to fix the remaining constants
and obtain the final result for the NS-limit of local $\IP^1 \times \IP^1$ at the orbifold
point.

We have to send $\hbar \rightarrow 2 \ri \hbar $ and introduce and overall factor of $1/8$ in order to compare the results to the Conifold computation.

This results seem to disagree with the results of \cite{Choi:2012jz}, but they
actually agree, except for the contribution coming from the constants of integration.
However, due to the change \eqref{eqn:stoS} the expressions change drastically.
The raw data can be found in appendix \ref{app:F0_orbifold_point}.
%
%
%
%
%
\subsection{\texorpdfstring{$\CO(-3) \rightarrow \IP^2$}{O(-3) -> P2} }
\label{sec:localP2}
\begin{figure}[h]
\begin{center}
\includegraphics[scale=0.18]{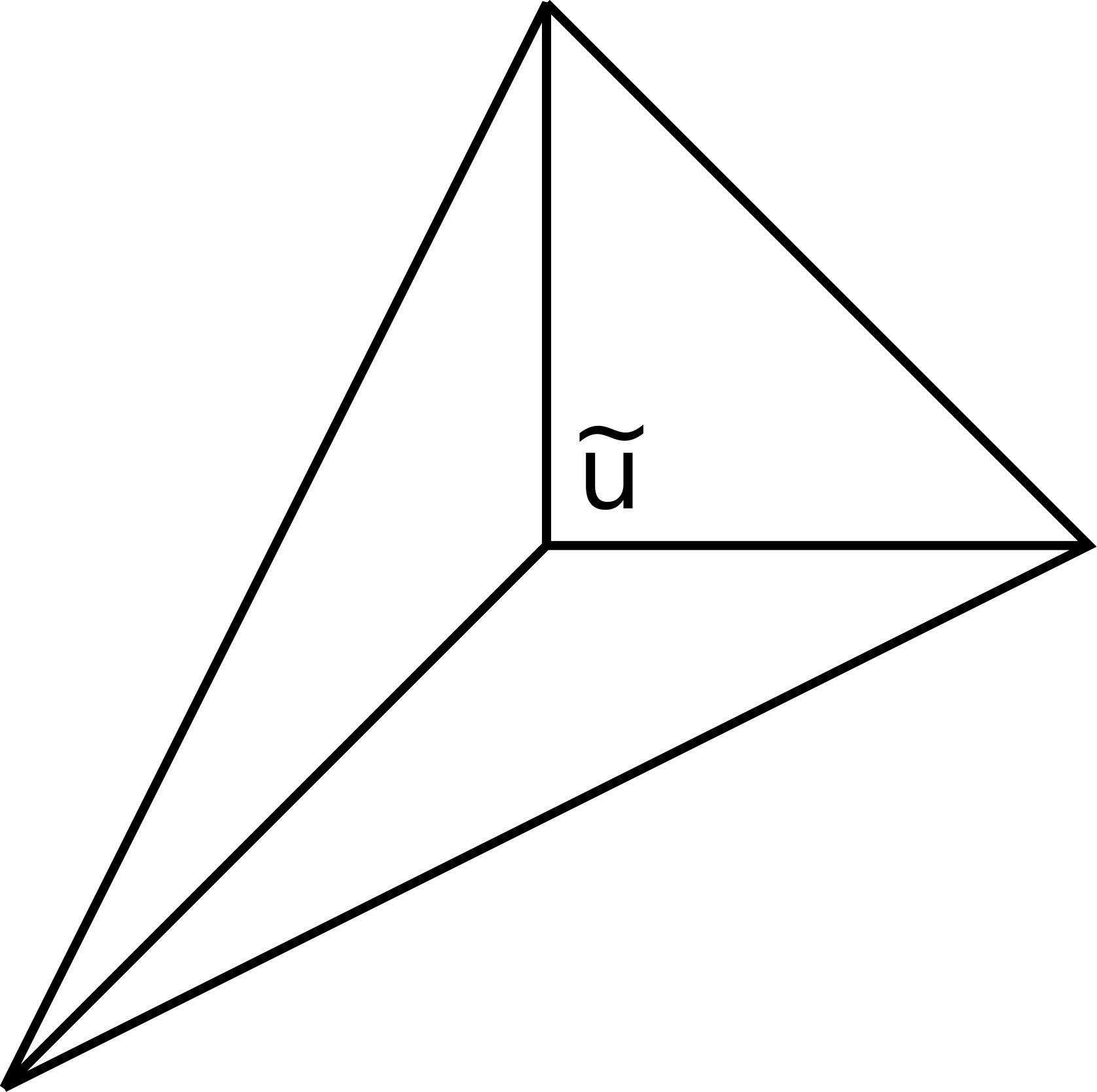}
\caption{\label{fig:poly1}Toric diagram of local $\CO(-3)\rightarrow \IP^2$.}
\end{center}
\end{figure}
\begin{equation} 
 \label{datap2} 
 \begin{array}{c|crrr|rl|} 
    \multicolumn{5}{c}{\nu_i }    &l^{(1)}&    \\ 
    D_u    &&     1&     0&   0&         -3&        \\ 
    D_1    &&     1&     1&   0&         1&         \\ 
    D_2    &&     1&     0&   1&         1&          \\ 
    D_3    &&     1&    -1&   -1&         1&           \\ 
  \end{array} \, . 
\end{equation}
The toric diagram of local $\IP^2$ is given in fig.~(\ref{fig:poly1}). By use of the toric data given in eg.~(\ref{datap2}) we find 
\begin{equation}
z=\frac{1}{\tilde{u}^3} \, .
\end{equation}
By defining $u=\frac{1}{\tilde{u}^3}$ we find for the quantum mirror curve
\begin{equation}
  -1 + e^x + e^p + u e^{\hbar/2} e^{-x} e^{-p} = 0 \, .
\end{equation}
The corresponding Schr\"odinger equation reads
\begin{equation}
  ( -1 + e^x  ) \psi(x) + \psi(x + \hbar)  + u e^{\hbar/2} e^{-x} \psi(x - \hbar) = 0 \, .
\end{equation}
Because we are not able to compute the full B-periods by only considering the patch
given by this parameterization, we have to use a second one, given by
\begin{equation}
 -1 + u e^{-x} + e^{-p} + e^{\hbar/2} e^x e^p = 0 \, .
\end{equation}
Therefore it is more convenient to use the operator approach in this case.
In \cite{Huang:2012kn} operators have been derived which enable us to write
higher order corrections to the periods in terms of the zero order period.
For local $\IP^2$ up to order four these are
\begin{subequations}
\begin{align}
  \CDD_2& = \frac{ \Theta_u^2  }{ 8 } \\
  \CDD_4 & = \frac{2 u(999 u - 5) \Theta_u  + 3 u (2619 u - 29)\Theta_u^2 }{640 \Delta^2} \label{eqn:P2_ops} 
\end{align}
\end{subequations}
where $\Delta = 1 + 27 u$.
The data for the elliptic curve is 
\begin{subequations}
\begin{align}
  g_2 & = 27 u^4 \left(24 u^3+1\right) \\
  g_3 & = 27 u^6 \left(216 u^6+36 u^3+1\right)
\end{align}
\end{subequations}
and from this together with the operators the $A$-period 
\begin{subequations}
\begin{equation}
  a = \log(u) -2 u^3 +15 u^6 -\frac{560 u^9}{3} 
  + \hbar^2 \left( -\frac{u^3}{4} +\frac{15 u^6}{2} -210 u^9 \right) + \CO(\hbar^4, u^{12})
\end{equation}
and the $B$-period
\begin{equation}
 a_D = -9 \left( \frac{1}{2} \log^2 u + \log(u) a 
  -u^3 +\frac{47 u^6}{4} -\frac{1486 u^9}{9} \right) + \CO(\hbar^2, u^{12})
\end{equation}
\end{subequations}
follow.
Having found these, we can easily integrate the special geometry relations to yield
\begin{subequations}
\begin{align}
 \CW^{(0)} & = 3 Q - \frac{45}{8} Q^2 + \frac{244}{9} Q^3 - \frac{12\ts{}333}{64} Q^4 + \frac{211\ts{}878}{125} Q^5 + \CO(Q^6) \\
 \CW^{(1)} & = - \frac{7}{8} Q + \frac{129}{16} Q^2 - \frac{589}{6} Q^3 + \frac{43\ts{}009}{32} Q^4 - \frac{392\ts{}691}{20} Q^5 + \CO(Q^6) \\
 \CW^{(2)} & = \frac{29 Q}{640}-\frac{207 Q^2}{64}+\frac{18447 Q^3}{160}-\frac{526859 Q^4}{160} +\frac{5385429 Q^5}{64}+ \CO(Q^6)
 \, .
\end{align}
\end{subequations}
\subsubsection{Orbifold point}
The orbifold point is given by the change of coordinates $\psi = - \frac{1}{3u^{1/3}}$,
which changes the logarithmic derivative of the large radius coordinate as
\begin{equation}
  \Theta_u \rightarrow - \frac{1}{3} \psi \partial_\psi = -\frac{1}{3}\Theta_\psi\,.
\end{equation}
According to \cite{Haghighat:2008gw} the classical part of the periods at this point can be written as
\begin{equation}
 \Pi_{\text{orb}} = \begin{pmatrix}
  \sigma_D \\ \sigma \\ 1 
 \end{pmatrix} =
 \begin{pmatrix}
   - 3 \partial_\sigma F_0^{\text{orb}} \\
   \sigma \\ 1
 \end{pmatrix} =
 \begin{pmatrix}
    B_2 \\ B_1 \\ 1
 \end{pmatrix}
\end{equation}
where 
\begin{equation}
  B_k = (-1)^{\frac{k}{3} + k + 1} \frac{(3 \psi)^k}{k} \sum_{n=0}^\infty \frac{\left[ \frac{k}{3} \right]^3_n}{\prod_{i=1}^3 \left[ \frac{k + i}{3} \right]_n} \psi^{3 n} \label{eqn:P2_orb_exp} \, .
\end{equation}
Here $[ a ]_n = a (a + 1) \ldots (a + n + 1)$ is the Pochhammer symbol.
Knowing the operators \eqref{eqn:P2_ops} and \eqref{eqn:P2_orb_exp}, computing the periods $\sigma$ and $\sigma_D$ to higher orders is very easy. 
We only have to change the coordinates of the operators to $\psi$ and apply them to
\eqref{eqn:P2_orb_exp}.

\ref{app:P2_orbifold}
The quantum periods are defined by the expansion
\begin{equation}
  t = \sigma = \sum_{i} \hbar^{2 i} \sigma^{2 i} \label{eqn:quantum_sigma} \\
  \quad\text{and}\quad
  \sigma_D = \sum_{i} \hbar^{2 i} \sigma^{2 i}_D  \,.
\end{equation}
The first few orders can be found in \eqref{eqn:P2_orb_a} for the $A$-period
and in \eqref{eqn:P2_orb_aD} for the $B$-period.
Inverting $t(\psi)$ and plugging it into $a_D$ gives us
\begin{equation}
  \sigma_D = \frac{\partial F^\text{NS}_\text{orb}}{\partial \sigma}
\end{equation}
By integrating this with respect to $\sigma$, we finally find the free energies
\begin{subequations}
\begin{align}
 F^{(0,0)} & = c_0 + \frac{1}{18} t^3 - \frac{1}{19\ts{}440} t^6 + \frac{1}{3\ts{}265\ts{}920} t^9 - \frac{1093}{349\ts{}192\ts{}166\ts{}400} t^{12} + 
 \frac{119\ts{}401}{2\ts{}859\ts{}883\ts{}842\ts{}816\ts{}000} t^{15} \\
 F^{(1,0)} & = c_1 + \frac{1}{648}t^3 - \frac{1}{46\ts{}656} t^6 + \frac{1319}{3\ts{}174\ts{}474\ts{}240} t^9 - \frac{10453}{1\ts{}142\ts{}810\ts{}726\ts{}400} t^{12} \nonumber \\
&\phantom{= c_1 + \frac{1}{648}t^3 - \frac{1}{46\ts{}656} t^6 + \frac{1319}{3\ts{}174\ts{}474\ts{}240} t^9 - }{}  + \frac{2\ts{}662\ts{}883}{12\ts{}354\ts{}698\ts{}200\ts{}965\ts{}120} t^{15}  \\
 F^{(2,0)} & = c_2 + \frac{1}{6480} t^3 - \frac{79}{8\ts{}398\ts{}080} t^6  + \frac{29}{65\ts{}318\ts{}400} t^9 - \frac{423\ts{}341}{22\ts{}856\ts{}214\ts{}528\ts{}000} t^{12} \nonumber \\
&\phantom{= c_2 + \frac{1}{6480} t^3 - \frac{79}{8\ts{}398\ts{}080} t^6  + \frac{29}{65\ts{}318\ts{}400} t^9 - }{} + \frac{1\ts{}332\ts{}163\ts{}447}{1\ts{}853\ts{}204\ts{}730\ts{}144\ts{}768\ts{}000} t^{15} \, .
\end{align}
\end{subequations}
Checking this against the results found in \cite{Huang:2011qx}, we find an exact agreement up to the constants of integration.
\subsubsection{Conifold point}
In order to find the free energies at the conifold, we have to solve the
Picard-Fuchs system at small $\Delta$, which is defined in terms
of the large radius variable by
\begin{equation}
  u = \frac{\Delta - 1}{27}\,,
\end{equation}
which changes the logarithmic derivative to
\begin{equation}
  \theta_u \rightarrow \theta_\Delta = (\Delta - 1)\partial_\Delta\,.
\end{equation}
The quantum corrections will be computed by making a coordinate transformation
to $\Delta$ in  \eqref{eqn:P2_ops}.
\begin{equation}
  \Pi = \begin{pmatrix}
  a t_c \\ 3 a {t_c}_D \\ 1
  \end{pmatrix},\quad\text{where}\quad a=-\frac{\sqrt{3}}{2\pi } \,.
\end{equation}
The flat coordinates with quantum corrections are given in terms of $\Delta$
by equations \eqref{eqn:P2_con_a} in appendix
\ref{app:P2_conifold}.
Plugging this into the $B$-periods given in \eqref{eqn:P2_con_aD} and integrating,
we finally arrive at the free energies
\begin{subequations}
\begin{align}
 F^{(0,0)} & = 
 c_0 + \frac{ a_0 t_c}{3} + t_c^2 \left(\frac{a_1}{6}+\frac{\log \left(t_c\right)}{6}-\frac{1}{12}\right)
 -\frac{t_c^3}{324} +\frac{t_c^4}{69984} +\frac{7\, t_c^5}{2361960}
 -\frac{529\, t_c^6}{1700611200} + \CO(t_c^7)
 \\
 F^{(1,0)} & = c_1 + \frac{\log \left(t_c \right)}{24}
 +\frac{7\, t_c}{432} -\frac{131\, t_c^2}{46656} -\frac{19\, t_c^3}{314928}
 +\frac{439\, t_c^4}{50388480} -\frac{1153\, t_c^5}{1530550080}
 + \CO(t_c^6) \\
 F^{(2,0)} & = 
 -\frac{7}{1920 t_c^2} + c_2 
 +\frac{1169 t_c}{12597120}
  -\frac{7367 t_c^2}{335923200}
 +\frac{16153 t_c^3}{6122200320}
 +\frac{7729 t_c^4}{881596846080}
+ \CO(t_c^5) 
\end{align}
\end{subequations}
where
\begin{equation}
 a_0=-\frac{\pi}{3} -1.678699904 \ri=
\frac{1}{i\sqrt{3}\Gamma\left(\frac{1}{3}\right)\Gamma\left(\frac{2}{3}\right)}G^{3\,3}_{2\,2}
\left({{\frac{1}{3}\ \frac{2}{3} \ 1} \atop {0 \ 0 \ 0}}\biggr| -1\right)
\quad\text{and}\quad
a_1 = \frac{3 \log(3)+1}{2\pi \ri}.
\end{equation}
This again matches the results given in \cite{Huang:2011qx} up to misprints and constants of integration.
%
%
%
%
%
%
\subsection{\texorpdfstring{local $\IF_1$}{local F1} }
\label{sec:localF1}
\begin{figure}[h]
\begin{center}
\includegraphics[scale=0.18]{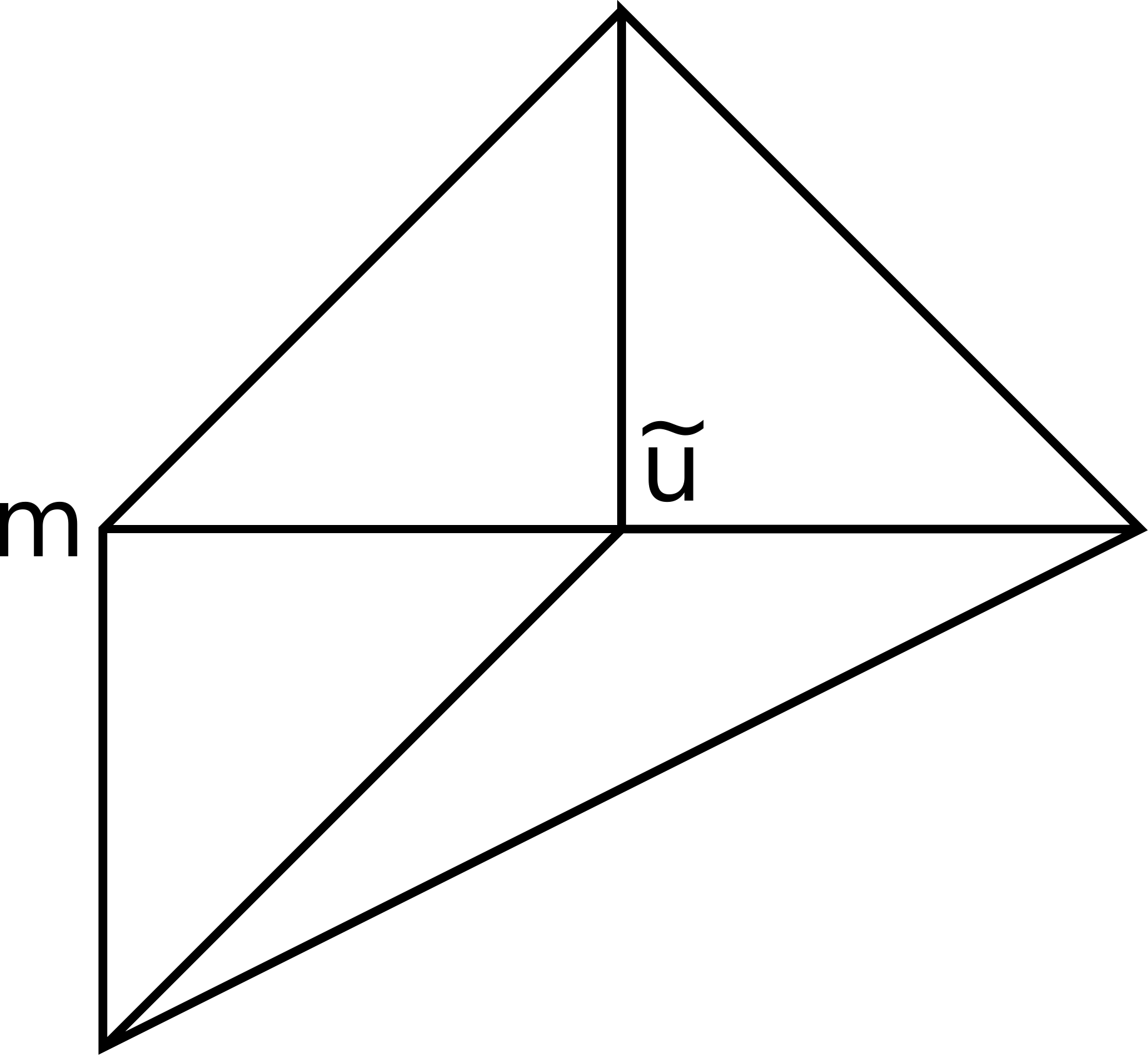}
\caption{\label{fig:poly3}Toric diagram of local $\mathds{F}_1$.}
\end{center}
\end{figure}
\begin{equation} 
 \label{dataf1} 
 \begin{array}{cc|crr|rrl|} 
    \multicolumn{5}{c}{\nu_i }    &l^{(1)}=l^{(f)} & l^{(2)}=l^{(b)}&\\ 
    D_u    &&     1&     0&   0&           -2& -1&     \\ 
    D_1    &&     1&     1&   0&            1&  0& \\ 
    D_2    &&     1&     0&   1&            0&  1& \\ 
    D_3    &&     1&    -1&   0&            1& -1& \\ 
    D_4    &&     1&    -1&  -1&            0&  1& \\ 
  \end{array} \ . 
\end{equation} 
The toric data of eq.~\eqref{dataf1} together with the definition of the trivial $m$ parameter in fig.~\eqref{fig:poly3} leads to the Batyrev coordiantes
\begin{equation}
 z_1 = \frac{m}{\tilde{u}^2},\quad z_2 = \frac{1}{m \tilde{u}} \, .
\end{equation}
We define $\tilde{u} = \frac{1}{u}$ and get for the quantum mirror curve
\begin{equation}
  H(x,p) = -1 + e^x + m u^2 \, e^{-x} + e^p + e^{- \hbar/2} \frac{u}{m} \, e^x e^{-p} \, .
  \label{eqn:locF1_curve}
\end{equation}
The coefficients of the classical Weierstrass normal form are
\begin{subequations}
\begin{align}
g_2(u, m) &= 27 u^4 (1 - 8 m u^2 + 24 u^3 + 16 m^2 u^4) \, ,\\
g_3(u, m) &= 27 u^6 (1 - 12 m u^2 + 36 u^3 + 48 m^2 u^4 - 144 m u^5 + 216 u^6 - 64 m^3 u^6) \, .
\end{align}
\end{subequations}
\subsubsection{Operator approach}
\label{sec:localF1_operator}
The curve \eqref{eqn:locF1_curve} has the following solution at zeroth order
\begin{align}
  S'_0 (x) & = \log \left(\frac{1}{2} e^{-x} \left(e^x-e^{2 x}-z_1 -\sqrt{\left(-e^x+e^{2 x}+z_1\right)^2+4 e^{3 x} z_2}  \right)\right) \, .
\end{align}
Further solutions can be found in appendix \ref{app:localF1}.\\
By partially integrating we find for the first operator
\begin{equation}
  \CDD_2 = \frac{m u^2\left(4m-9u\right)}{6 \delta }\Theta_u + \frac{4m-3u-16m^2 u^2+36u^3 m}{24\delta}\Theta_u^2 \, ,
  \label{eqn:F1_HOp}  
\end{equation}
with $\delta = \left(-8m+9u\right)$.
Higher order operators can as well be found in appendix \ref{app:localF1}.
Calculation of the nontrivial quantum periods leads to
\begin{subequations}
\begin{align}
a &= \log(u) + m u^2 - 2 u^3 - \frac{1}{4}u^3 \hbar^2 + \CO(\hbar^4, u^4)\, ,\\
a_D &= - 4 \log(u)^2  - \log(u) \log(m) - \log(u^8 m) \left ( m u^2 - 2 u^3 - \frac{1}{4} u^3 \hbar^2 \right) + \frac{u}{m} + u^2( \frac{1}{4m^2} -4m) \nonumber \\
& \phantom{={}}  + 10 u^3 + \frac{1}{9 m^3}u^3 - \frac{1}{24} \hbar ^2 \left (4 + \frac{u}{m} + u^2( \frac{1}{m^2} +8m) +u^3(\frac{1}{m^3}-62) \right)+ \CO(\hbar^4, u^4)\, .
\end{align}
\end{subequations}
With the nontrivial A-period we find for the mirror map
\begin{equation}
u(Q_u) = Q_u - m Q_u^3 + 2 Q_u^4 + \frac{1}{4}Q_u^4 \hbar^2 + \CO(\hbar^4, Q_u^5)\, .
\end{equation}
The nontrivial coordinate $Q_u$ and the trivial parameter $m$ can be translated back to the usual description via two logarthmic solutions of the Picard-Fuchs equations. The connection is given by
\begin{equation}
Q_u= Q_1^{1/3} Q_2^{1/3} ,\; m=Q_1^{1/3} Q_2^{-2/3} \, .
\end{equation}
These relations can be checked perturbativly. Using these relations after integrating the special geometry relation we find the instanton numbers in the NS limit listed in the tables \ref{tab:instf1genus0}, \ref{tab:instf1genus1}, \ref{tab:instf1genus2} and \ref{tab:instf1genus3}.
\begin{table}[h!]
\begin{center}
\begin{tabular}{c|cccccc} 
 & $d_1$ & 0 & 1 & 2 & 3 & 4  \\
\hline
 $d_2$ & & & & & &  \\
0 &  &   & $1$ &   &   &  \\
1 &  & $-2$ & $3$ &   &   &  \\
2 &  &   & $5$ & $-6$ &   &  \\
3 &  &   & $7$ & $-32$ & $27$ &  \\
4 &  &   & $9$ & $-110$ & $286$ & $-192$\\
\end{tabular}
\caption{\label{tab:instf1genus0} The instanton numbers for local $\mathds{F}_1$ at order $\hbar^0$.}
\end{center}
\end{table}
\begin{table}[h!]
\begin{center}
\begin{tabular}{c|cccccc} 
 & $d_1$ & 0 & 1 & 2 & 3 & 4  \\
\hline
 $d_2$ & & & & & &  \\
0 &  &   &   &   &   &  \\
1 &  & $-1$ & $4$ &   &   &  \\
2 &  &   & $20$ & $-35$ &   &  \\
3 &  &   & $56$ & $-368$ & $396$ &  \\
4 &  &   & $120$ & $-2055$ & $6732$ & $-5392$\\
\end{tabular}
\caption{\label{tab:instf1genus1} The instanton numbers for local $\mathds{F}_1$ at order $\hbar^2$.}
\end{center}
\end{table}
\begin{table}[h!]
\begin{center}
\begin{tabular}{c|cccccc} 
 & $d_1$ & 0 & 1 & 2 & 3 & 4  \\
\hline
 $d_2$ & & & & & &  \\
0 &  &   &   &   &   &  \\
1 &  &   & $1$ &   &   &  \\
2 &  &   & $21$ & $-56$ &   &  \\
3 &  &   & $126$ & $-1352$ & $1875$ &  \\
4 &  &   & $462$ & $-12892$ & $55363$ & $-53028$\\
\end{tabular}
\caption{\label{tab:instf1genus2} The instanton numbers for local $\mathds{F}_1$ at order $\hbar^4$.}
\end{center}
\end{table}
\begin{table}[h!]
\begin{center}
\begin{tabular}{c|cccccc} 
 & $d_1$ & 0 & 1 & 2 & 3 & 4  \\
\hline
 $d_2$ & & & & & &  \\
0 &  &   &   &   &   &  \\
1 &  &   &   &   &   &  \\
2 &  &   & $8$ & $-36$ &   &  \\
3 &  &   & $120$ & $-2412$ & $4344$ &  \\
4 &  &   & $792$ & $-41594$ & $242264$ & $-277430$\\
\end{tabular}
\caption{\label{tab:instf1genus3} The instanton numbers for local $\mathds{F}_1$ at order $\hbar^6$.}
\end{center}
\end{table}
\subsubsection{Difference equation}
\label{sec:F1_diff_eq}
Another way to handle this problem would be to extract the relevant data
directly from the integrals over the quantum differentials. For the A-periods
this is quite straightforward and the non-logarithmic part is just given by 
the residues of the expansion around the large radius coordinates. 
The computation of the B-periods though is not as straightforward. 
We have to change the parameterization of the curve in order to
find all contributions. Unfortunately we are not certain about
the way to systematically find these parameterizations.
The following curves yield all the parts needed for getting the
correct B-period, at least for zeroth order in $\hbar$. 
\begin{align}
  \mathcal{A}: &  -1 + e^p + e^x + u^2 m\, e^{-x} - e^{\hbar/2} u^3\, e^{-p-x} \nonumber \\
  \mathcal{B}: &  -1 + e^p + e^x + u^2 m\, e^{-p} - e^{\hbar/2} u^3\, e^{-p - x} \nonumber  \\
  \mathcal{C}: &  -1 + e^p + e^x + u^2 m\, e^{-x} - e^{-\hbar/2} \frac{u}{m} e^{-p + x} \, .
\end{align}
The A-period is
\begin{subequations}
\begin{equation}
 a = \log(u) + m u^2 - \left(\sqrt{q}+\frac{1}{\sqrt{q}}\right) u^3 
 + \frac{3 m^2 u^4}{2} + \CO(u^5)
\end{equation}
while the B-period, after summing up the contributions from $\mathcal{A},\mathcal{B}$ and $\mathcal{C}$ is given by
\begin{equation}
  a_D = -\log (u) \log \left(m u^4\right) - \log \left(m u^8\right) \tilde{a} + \tilde{a}_D
\end{equation}
\end{subequations}
where
\begin{align}
 \tilde{a}_D &= \frac{\sqrt{q} u \log (q)}{m (q-1)} + 
 \frac{u^2 \left(q-4 m^3 (q+1)^2\right) \log (q)}{2 m^2 \left(q^2-1\right)} \nonumber \\
&\phantom{= \frac{\sqrt{q} u \log (q)}{m (q-1)} + }{} +
 \frac{u^3 \left(6 m^3 \left(2 q^4+3 q^3+5 q^2+3 q+2\right)+q^2\right) \log (q)}{3 m^3 \sqrt{q} \left(q^3-1\right)} + \CO(u^4)
\end{align}
which can be pieced together from the contributions of the different parameterizations 
like
\begin{equation}
 \tilde{a}_D = -3 I_\mathcal{A} -4 I_\mathcal{B} - I_\mathcal{C}
\end{equation}
after symmetrization with respect to $\hbar \rightarrow -\hbar$, in order to
get rid of the odd sector in $\hbar$.

By integrating the B-period with respect to $Q_t$ we finally find the
free energy
\begin{equation}
 \CW = \frac{\left( (q + 1)Q_1 -\sqrt{q} Q_2 \right) }{1-q}
 + \frac{\left(q^2+1\right) Q_1^2 - q Q_2^2}{4 \left(1 - q^2\right)}
 -  \frac{\left(4 q^2+q+4\right) Q_1 Q_2}{3 (1-q) \sqrt{q}}
 + \CO(Q_i^3) \,.
\end{equation}
\subsection{\texorpdfstring{$\mathcal{O}(-K_{\mathds{F}_2})\rightarrow \mathds{F}_2$}{O(-KF2) -> F2}}
\label{sec:localF2}
\begin{figure}[h]
  \centering
  \includegraphics[scale=0.6]{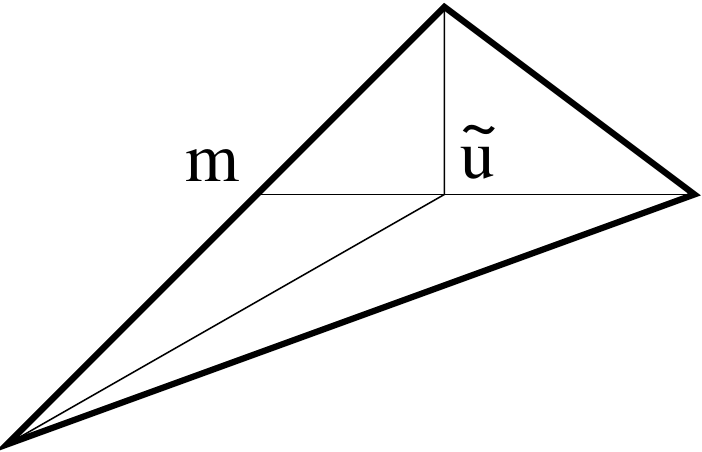}
  \caption{Toric diagram of $\mathcal{O}(-K_{\mathds{F}_2})\rightarrow \mathds{F}_2$.}
  \label{fig:f2}
\end{figure}
\begin{equation} 
 \label{dataf2} 
 \begin{array}{c|crr|rrrl|} 
    \multicolumn{5}{c}{\nu_i }    &l^{(1)}=l^{(f)}&l^{(2)}=l^{(b)}&\\ 
    D_u    &&     1&     0&   0&          -2& 0&\\ 
    D_1    &&     1&     1&   0&           1& 0&\\ 
    D_2    &&     1&     0&   1&           0& 1&\\ 
    D_5    &&     1&    -1&   0&           1&-2&\\ 
    D_6    &&     1&    -2&   -1&          0& 1& \\ 
  \end{array} \, . 
\end{equation} 
This geometry is denoted in \cite{Huang:2013yta} as polyhedron $4$. The toric diagram is depicted fig.~(\ref{fig:f2}). The toric data is given in eq.~(\ref{dataf2}). With the toric data we find for the moduli
\begin{equation}
 z_1=\frac{m}{\tilde{u}^2},\; z_2=\frac{1}{m^2}\, .
\end{equation}
With the definition $\tilde{u}^2=\frac{1}{u}$ we find that the elliptic mirror curve does not have any quantum corrections and looks like 
\begin{equation}
	H=1+e^x+e^p+m u e^{2x}+\frac{1}{m^2} e^{-p} \, .
\end{equation}
The coefficients of the classical Weierstrass normal form are
\begin{subequations}
\begin{align}
g_2(u, m) &= 27 u^4 \left((1-4 m u)^2-48 u^2\right) \, ,\\
g_3(u, m) &=-27 u^6 \left(64 m^3 u^3-48 m^2 u^2-288 m u^3+12 m u+72 u^2-1\right) 
 \, .
\end{align}
\end{subequations}
The zeroth order solution to the resulting Schr\"odinger equation is
\begin{equation}
  S_0^\prime (x)= \log \left (\frac{1}{2} \left (-1-e^x-e^{2 x} m u -\sqrt{(1+e^x+e^{2 x} m u^2)^2-\frac{4}{m^2}}\right ) \right ) \, .
\end{equation}
Some higher order WKB functions can be found in appendix \ref{app:localF2}.\\
Partially integrating the WKB functions we find for operators mapping the zeroth order periods to higher periods
\begin{subequations}
\begin{align}
  \CDD_2 &=-\frac{1}{6} (m u) \Theta_u + \frac{1}{12} (1-4 m u)\Theta_u^2 \, ,\\
  \CDD_4 &=\frac{u^2}{180 \Delta^2}  \left(-4 m \left(3 m^2+28\right) u+m^2-64 m \left(m^4-92 m^2+352\right) u^3 \right. \nonumber \\
  & \phantom{=\frac{u^2}{180 \Delta^2}  ( } {} \left. +16 \left(3 m^4-94 m^2+552\right) u^2+30\right) \Theta_u \nonumber\\
& \phantom{={}} {}+ \frac{u}{360 \Delta^2} \left(-96 m \left(m^2+5\right) u^2+4 \left(4 m^2+61\right) u-256 m \left(m^4-92 m^2+352\right) u^4 \right. \nonumber \\
&\phantom{=+ \frac{u}{360 \Delta^2} ( } {} \left. +64 \left(4 m^4-123 m^2+652\right) u^3-m\right) \Theta_u^2 \, .
  \label{eqn:f2_HOp}
\end{align}
\end{subequations}
where
\begin{equation}
 \Delta = 16 m^2 u^2-8 m u-64 u^2+1\,.
\end{equation}
Higher order operators can as well be found in the appendix \ref{app:localF2}.\\
This geometry has a particularly interesting property, namely that one can calculate more than one A-period by taking the residue at another point. This allows an additional check of the operators.\\
Proceeding with the calculation of the nontrivial periods leads to
\begin{subequations}
\begin{align}
a &= \log (u) + mu^2 - 2 m u^3  +\frac{1}{6}mu^2 \hbar^2 -\frac{3}{2}mu^3\hbar^2 + \CO(\hbar^4, u^4) \, ,\\
a_D &=  -\frac{5}{2} \log (u)^2 + \log (u) \left ( - 5 m u^2 + 10 m u^3 - 
  \frac{5}{6} m u^2 \hbar^2 + \frac{15}{2} m u^3 \hbar^2 \right ) + u + \frac{u^2}{4} - 4 m u^2 + \frac{4}{9} u^3 \nonumber \\
&\phantom{=} + 9 m u^3 + \frac{1}{3}m^3 u^3 + \left ( -\frac{5}{12} + \frac{u}{12} + \frac{u^2}{12} - \frac{4}{3} m u^2 + \frac{1}{3}u^3 + \frac{45}{4} m u^3 + \frac{1}{4} m^3 u^3 \right ) \hbar ^2\, + \CO(\hbar^4, u^4).
\end{align}
\end{subequations}
After exponentiating the A-period we find for the mirror map
\begin{equation}
u(Q_u)=Q_u - m Q_u^3 - 3 Q_u^5 + m^2 Q_u^5 - \hbar^2 Q_u^5 - \frac{1}{12} \hbar^4 Q_u^5 + \CO(\hbar^6, Q_u^6)\, .
\end{equation}
In the following we use the relation
\begin{equation}
  \frac{1+Q_2}{\sqrt{Q_2}}=\frac{1}{\sqrt{z_2}}=m \, ,
\end{equation}
which does not get any quantum corrections and is thus like a trivial period. Additionally we use $Q_u=Q_1^{1/2} Q_2^{1/4}$ to find after integrating the special geometry relation the instanton numbers listed in the tables \ref{tab:bpsf2genus0}, \ref{tab:bpsf2genus1}, \ref{tab:bpsf2genus2} and \ref{tab:bpsf2genus3}.

Notice that we see at least a discrepancy in the contribution $\hat{n}_n^{0,1} = 1$ when
comparing this to results from the (refined) topological vertex. Something along
those lines has also been mentioned in \cite{Chiang:1999tz}.
\begin{table}[h!]
\begin{center}
\begin{tabular}{c|cccccc} 
 & $d_1$ & 0 & 1 & 2 & 3 & 4 \\
\hline
 $d_2$ & & & & & & \\
0 &  &   &   &   &   &   \\
1 &  & $2$ & $2$ &   &   &   \\
2 &  &   & $4$ &   &   &   \\
3 &  &   & $6$ & $6$ &   &   \\
4 &  &   & $8$ & $32$ & $8$ &   \\
\end{tabular}
\caption{\label{tab:bpsf2genus0} The instanton numbers for local $\mathds{F}_2$ at order $\hbar^0$..}
\end{center}
\end{table}
\begin{table}[h!]
\begin{center}
\begin{tabular}{c|cccccc} 
 & $d_1$ & 0 & 1 & 2 & 3 & 4 \\
\hline
 $d_2$ & & & & & & \\
0 &  &   &   &   &   &   \\
1 &  & $1$ & $1$ &   &   &   \\
2 &  &   & $10$ &   &   &   \\
3 &  &   & $35$ & $35$ &   &   \\
4 &  &   & $84$ & $368$ & $84$ &   \\
\end{tabular}
\caption{\label{tab:bpsf2genus1} The instanton numbers for local $\mathds{F}_2$ at order $\hbar^2$.}
\end{center}
\end{table}
\begin{table}[h!]
\begin{center}
\begin{tabular}{c|cccccc} 
 & $d_1$ & 0 & 1 & 2 & 3 & 4 \\
\hline
 $d_2$ & & & & & & \\
0 &  &   &   &   &   &   \\
1 &  &   &   &   &   &   \\
2 &  &   & $6$ &   &   &   \\
3 &  &   & $56$ & $56$ &   &   \\
4 &  &   & $252$ & $1352$ & $252$ &   \\
\end{tabular}
\caption{\label{tab:bpsf2genus2} The instanton numbers for local $\mathds{F}_2$ at order $\hbar^4$.}
\end{center}
\end{table}
\begin{table}[h!]
\begin{center}
\begin{tabular}{c|cccccc} 
 & $d_1$ & 0 & 1 & 2 & 3 & 4 \\
\hline
 $d_2$ & & & & & & \\
0 &  &   &   &   &   &   \\
1 &  &   &   &   &   &   \\
2 &  &   & $1$ &   &   &   \\
3 &  &   & $36$ & $36$ &   &  \\
4 &  &   & $330$ & $2412$ & $330$ &   \\
\end{tabular}
\caption{\label{tab:bpsf2genus3} The instanton numbers for local $\mathds{F}_2$ at order $\hbar^6$.}
\end{center}
\end{table}
\subsection{\texorpdfstring{$\mathcal{O}(-K_{\mathcal{B}_2})\rightarrow \mathcal{B}_2$}{O(-KB2) -> B2}}
\label{sec:localB2}
\begin{figure}[h]
\begin{center}
  \includegraphics[scale=0.2]{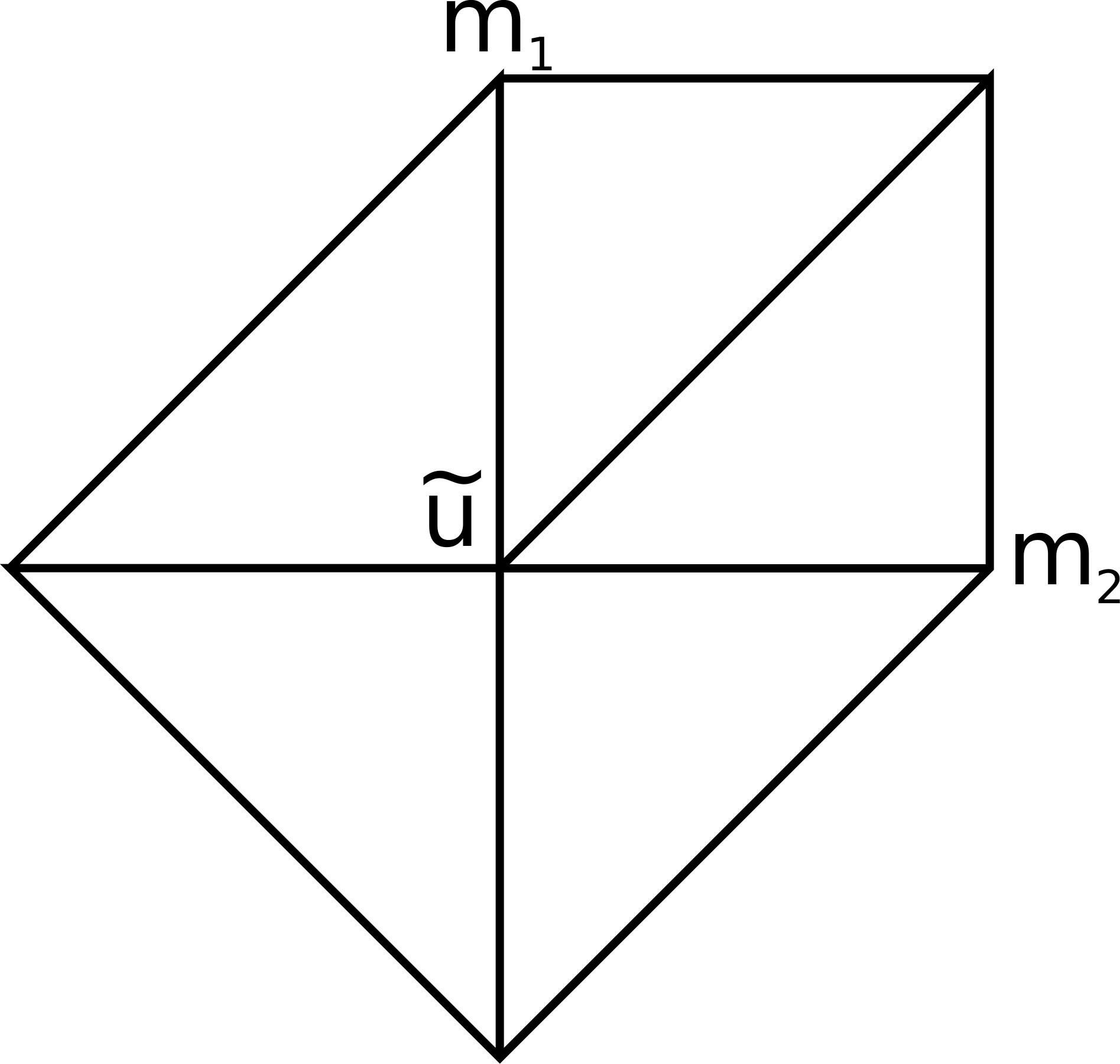}
  \caption{\label{fig:poly5}Toric diagram of $\mathcal{O}(-K_{\mathcal{B}_2})\rightarrow \mathcal{B}_2$ with the assigned mass parameters and the modulus $\tilde{u}$.}
\end{center}
\end{figure}
The toric diagramm of this geometry is given in fig.~(\ref{fig:poly5}). Its toric data is given in eq.~(\ref{datadp2}).
\begin{equation} 
 \label{datadp2} 
 \begin{array}{cc|crr|rrrl|} 
    \multicolumn{5}{c}{\nu_i }    &l^{(1)} & l^{(2)} &l^{(3)}&\\ 
    D_u    &&     1&     0&   0&           -1& -1& -1&     \\ 
    D_1    &&     1&     1&   0&           -1&  1&  0&\\ 
    D_2    &&     1&     1&   1&            1& -1&  1&\\ 
    D_3    &&     1&     0&   1&            0&  1& -1&\\ 
    D_4    &&     1&    -1&   0&            0&  0&  1&\\ 
    D_5    &&     1&     0&  -1&            1&  0&  0&
  \end{array} \ . 
\end{equation}
With the toric data we find for the moduli
\begin{equation}
   z_1 = \frac{1}{\tilde{u} m_2},\; z_2 = \frac{m_1 m_2}{\tilde{u}},\; z_3 = \frac{1}{\tilde{u} m_1} \, .
\end{equation}
For the quantum mirror curve for the $\mathcal{B}_2$ geometry we have with $\tilde{u}=\frac{1}{u}$
\begin{equation}
  H= 1 + e^x + e^p + \frac{u}{m_2} e^{-\frac{\hbar}{2} + x - p} + u m_1 m_2 e^{-\frac{\hbar}{2} + p - x} + m_2 u^2 e^{-x}\, .
\end{equation}
The coefficients of the Weierstrass normal form are given by
\begin{subequations}
\begin{align}
  g_2 & = 27 u^4 ( \left(16 m_1^2-16 m_2 m_1+16 m_2^2\right) u^4+24 u^3-\left(8 m_1+8 m_2\right) u^2+1 ) \, ,\\
  g_3 & = 27 u^6 \left(1 -\left(12 m_1+12 m_2\right) u^2 +36 u^3 +\left(48 m_1^2+24 m_2 m_1+48 m_2^2\right) u^4 \right .\nonumber \\
  &\phantom{ ={}} \left . -\left(144 m_1+144 m_2\right) u^5 + \left(-64 m_1^3+96 m_2 m_1^2+96 m_2^2 m_1-64 m_2^3+216\right) u^6 \right) \, .
\end{align}
\end{subequations}
The zeroth order solution to the resulting Schr\"odinger equation is
\begin{equation}
  S_0^\prime (x)= \log\left (-\frac{e^{x}+e^{2 x}+u^2 m_2+\sqrt{-4 e^{2 x} \frac{u}{m_2} (e^x+u m_1 m_2)+(e^x+e^{2 x}+u^2 m_2)^2}}{2 (e^x+u m_1 m_2)}\right ) \, .
\end{equation}
Due to an additional pole in the higher order WKB functions stemming from the non-quadratic term in the root this case and the following are considerably more complicated than the geometries $\mathds{F}_0$, $\mathds{F}_1$ and $\mathds{F}_2$. Nontheless one finds the operator
\begin{align}
\CDD_2 & =-\frac{1}{6 \delta} \left (u^2 (m_2 (4 m_2 - 9 u) u + 4 m_1^3 m_2 u (m_2 + u) + m_1^2 (-5 m_2 + 4 u + 4 m_2^3 u - 16 m_2^2 u^2) \right . \nonumber \\
  & \phantom{ = {} } \left .+ m_1 (-5 m_2^2 + 20 m_2 u - 9 u^2 + 4 m_2^3 u^2))\right) \Theta_u \nonumber \\
  & \phantom{ = {} } + \frac{1}{24 \delta} \left (-16 m_1^3 m_2 u^3 (m_2 + u) + u (-3 u - 16 m_2^2 u^2 + 4 m_2 (1 + 9 u^3)) + 4 m_1^2 u (6 m_2 u - 4 u^2 \right .\nonumber \\
  & \phantom{ = {} } \left . - 4 m_2^3 u^2 + m_2^2 (1 + 16 u^3)) + m_1 (24 m_2^2 u^2 - 16 m_2^3 u^4 - m_2 (5 + 92 u^3) + 4 (u + 9 u^4))\right ) \Theta_u^2 \, ,
\end{align}
with $\delta=((8 m_2 - 9 u) u + 4 m_1^2 m_2 (2 m_2 - u) u + m_1 (-7 m_2 + 8 u - 4 m_2^2 u^2))$.
Notice that this operator simplifies to the operator, that maps the classical periods of local $\IF_1$ (section \ref{sec:localF1}) to the second order, if we take the limit $m_1 \rightarrow 0$ or $m_2 \rightarrow 0$, as we would expect from \eqref{datadp2}.
So we indeed find the correct amplitude when blowing down and passed this consistency check successfully.

Calculation of the nontrivial quantum periods leads to
\begin{subequations}
\begin{align}
  a &= \log(u) + m_1 u^2 + m_2 u^2 - 2 u^3 - \frac{u^3}{4} \hbar^2 + \CO(\hbar^4, u^4)\, , \\ 
  a_D &=-\frac{7}{2} \log(u)^2 - \log(-m_1 m_2) \log(u) + \frac{u}{m_1} + \frac{u}{m_2} + m_1 m_2 u \nonumber \\
& \phantom{={}} - \frac{1}{24}\left( 5+ \frac{u}{m_1} + \frac{u}{m_2} + m_1 m_2 u\right)\hbar^2   + \CO(\hbar^4, u^2)\, .
\end{align}
\end{subequations}
Using this we find for the mirror map
\begin{equation}
  u(Q_u) = Q_u - (m_1 + m_2) Q_u^3 + 2 Q_u^4 + m_1^2 Q_u^5 - m_1 m_2 Q_u^5 + m_2^2 Q_u^5 + (\frac{1}{4}Q_u^4 - m_1 m_2 Q_u^5) \hbar^2 + \CO(\hbar^4, Q_u^6)\, .
\end{equation}
Since here we do not have any points sitting on an edge we can invert the relations between $z$ and $u,m_i$ and find for $u,m_i$ in dependence of the coordinates $Q_i$
\begin{equation}
u=Q_1^{1/3} Q_2^{1/3} Q_3^{1/3},\; m_1=Q_1^{1/3}Q_2^{1/3}Q_3^{-2/3},\; m_2=Q_1^{-2/3}Q_2^{1/3}Q_3^{1/3} \, .
\end{equation}
After integrating the special geometry relation and plugging this in we find the following free energy
\begin{subequations}
\begin{align}
  F^{(0,0)}&= (Q_1+Q_2+Q_3) + \left (\frac{Q_1^2}{8}-2 Q_1 Q_2+\frac{Q_2^2}{8}-2 Q_2 Q_3+\frac{Q_3^2}{8} \right) \nonumber\\
& \phantom{={}} + \left (\frac{Q_1^3}{27}+\frac{Q_2^3}{27}+3 Q_1 Q_2 Q_3+\frac{Q_3^3}{27}\right)+ \left(\frac{Q_1^4}{64}-\frac{Q_1^2 Q_2^2}{4}+\frac{Q_2^4}{64}-4 Q_1 Q_2^2 Q_3-\frac{Q_2^2 Q_3^2}{4}+\frac{Q_3^4}{64}\right)\nonumber\\
& \phantom{={}} + \left(\frac{Q_1^5}{125}+\frac{Q_2^5}{125}+5 Q_1^2 Q_2^2 Q_3+5 Q_1 Q_2^2 Q_3^2+\frac{Q_3^5}{125}\right) + \CO(Q^6)\, ,\\
F^{(1,0)}&= \left(-\frac{Q_1}{24}-\frac{Q_2}{24}-\frac{Q_3}{24}\right)+\left(-\frac{Q_1^2}{48}-\frac{Q_1 Q_2}{6}-\frac{Q_2^2}{48}-\frac{Q_2 Q_3}{6}-\frac{Q_3^2}{48}\right)\nonumber\\
& \phantom{={}} +\left(-\frac{Q_1^3}{72}-\frac{Q_2^3}{72}+\frac{7 Q_1 Q_2 Q_3}{8}-\frac{Q_3^3}{72}\right)+\left(-\frac{Q_1^4}{96}-\frac{Q_1^2 Q_2^2}{12}-\frac{Q_2^4}{96}-\frac{7 Q_1 Q_2^2 Q_3}{3}-\frac{Q_2^2 Q_3^2}{12}-\frac{Q_3^4}{96}\right)\nonumber\\
& \phantom{={}} +\left(-\frac{Q_1^5}{120}-\frac{Q_2^5}{120}+\frac{115 Q_1^2 Q_2^2 Q_3}{24}+\frac{115 Q_1 Q_2^2 Q_3^2}{24}-\frac{Q_3^5}{120}\right) + \CO(Q^6)\, .
\end{align}
\end{subequations}
\subsection{local \texorpdfstring{$\CB_1(\IF_2)$}{CB1(F2)}}
\label{sec:poly6}
\begin{figure}[h]
  \centering
  \includegraphics[scale=0.2]{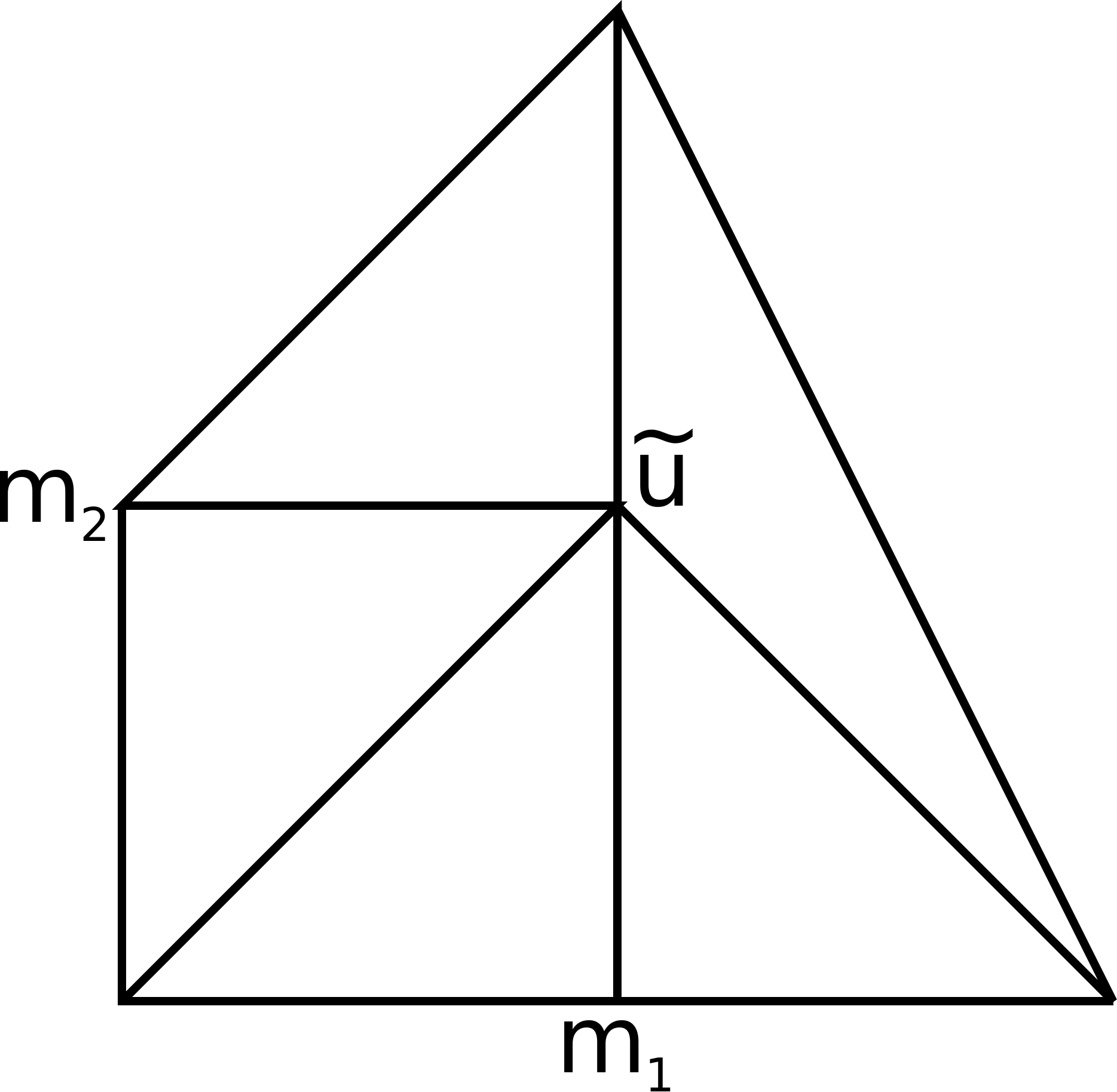}
  \caption{Toric diagram of $\CB_1(\IF_2)$ with the assigned masses.}
  \label{fig:of2}
\end{figure}
\begin{equation} 
 \label{dataof2} 
 \begin{array}{cc|crr|rrrl|} 
    \multicolumn{5}{c}{\nu_i }    &l^{(1)}&l^{(2)}&l^{(3)}&\\ 
    D_u    &&     1&     0&   0&          -1&-1& 0&\\ 
    D_1    &&     1&     1&  -1&           0& 0& 1&\\ 
    D_2    &&     1&     0&   1&           1& 0& 0&\\ 
    D_3    &&     1&    -1&   0&          -1& 1& 0&\\ 
    D_4    &&     1&    -1&  -1&           1&-1& 1&\\ 
    D_5    &&     1&     0&  -1&           0& 1&-2&\\ 
  \end{array} \ . 
\end{equation} 
The toric data of this geometry can be found in eq.~(\ref{dataof2}) and the toric diagramm with the used mass assignement is given in fig.~(\ref{fig:of2}). The toric data leads to the coordinates
\begin{equation}
  z_1=\frac{1}{\tilde{u} m_2}\, ,\quad z_2=\frac{m_1 m_2}{\tilde{u}}\, ,\quad z_3=\frac{1}{m_1^2} \, .
\end{equation}
By defining $u=\frac{1}{\tilde{u}}$ we find for the quantum curve
\begin{equation}
  H=1 + e^{x} + e^{p} + m_1 u^2 e^{2 p} + \frac{m_2}{m_1}u e^{-\frac{\hbar}{2} + p - x} + \frac{1}{m_1^2} e^{-x}\, .
\end{equation}
The coefficients of the Weierstrass normal form of this curve are given by
\begin{subequations}
\begin{align}
g_2(u) &= 27 u^4 (1 - 8 m_1 u^2 + 24 m_2 u^3 - 48 u^4 + 16 m_1^2 u^4) \, ,\\
g_3(u) &= 27 u^6 (1 - 12 m_1 u^2 + 36 m_2 u^3 - 72 u^4 + 48 m_1^2 u^4 - 144 m_1 m_2 u^5 \nonumber \\
& \phantom{={}} + 288 m_1 u^6 - 64 m_1^3 u^6 + 216 m_2^2 u^6)\, .
\end{align}
\end{subequations}
By solving the Schr\"odinger equation resulting from the quantum curve perturbatively in $\hbar$ we find for the zeroth order WKB function which is equivalent to the classical differential
\begin{equation}
  S_0^\prime (x)= \log \left (    \frac{-1 - e^{-x} \frac{m_2}{m_1} u - e^{-x} \sqrt{ -4 e^{x} m_1 u^2 (e^x + e^{2 x} + \frac{1}{m_1^2}) + (e^x + \frac{m_2}{m_1}u)^2}}{2 m_1 u^2} \right ) \, .
\end{equation}
For the operator mapping the zeroth order periods to the second order periods we find
\begin{align}
  \CDD_2 &= \frac{u^2 (4 m_1^2 m_2 u (-m_2 + u) - 12 m_2 u (m_2 + 2 u) - m_1 (5 m_2 + 4 u + 9 m_2^3 u^2))}{6 \delta} \Theta_u \nonumber \\
& \phantom{={}} + \frac{1}{24 \delta} \left (4 u - 16 m_1 u^3 - 4 m_2^2 (-m_1 u + 15 u^3 + 4 m_1^2 u^3) + m_2^3 (3 u^2 - 36 m_1 u^4)\right .\nonumber \\
& \phantom{={}} \left . + m_2 (5 - 24 m_1 u^2 - 96 u^4 + 16 m_1^2 u^4) \right )\Theta_u^2 \, ,
\label{eqn:OF2_HOp}
\end{align}
with $\delta= (8 u + 8 m_1 m_2^2 u + 9 m_2^3 u^2 + m_2 (7 + 4 m_1 u^2))$.
The three logarithmic solutions of the Picard-Fuchs equations yield only one nontrivial period. This can be seen by combining them in the following way
\begin{align}
\label{eq:relationsof2}
Q_u&=Q_1^{\frac{1}{2}} Q_2^{\frac{1}{2}} Q_3^{\frac{1}{4}}\, ,\\
m_1&=\frac{1+Q_3}{\sqrt{Q_3}}\, ,\\
m_2&=Q_1^{-\frac{1}{2}} Q_2^{\frac{1}{2}} Q_3^{\frac{1}{4}}
\end{align}
where $m_1$ is not just given by a simple linear combination of the periods. Using the operator we find for the nontrivial periods in the NS limit
\begin{subequations}
\begin{align}
a &= \log(u)+m_1 u^2+(-2 m_2-\frac{m_2 \hbar^2}{4}) u^3+\left (3+\frac{3}{2} m_1^2+\hbar^2 \right ) u^4\nonumber \\
& \phantom{={}} +\left (-12 m_1 m_2-\frac{7}{2} m_1 m_2 \hbar^2 \right ) u^5 + \CO(\hbar^4, u^6) \, ,\\
a_D &= - \frac{7}{2} \log(u)^2 - \log(-m_2) \log(u) - m_1 u^2 \log(-m_2 u^7) \nonumber \\
& \phantom{={}} + \frac{u}{m_2} + m_1 m_2 u -u^2 \left (4 m_1-\frac{1}{4 m_2^2} +\frac{m_2^2}{2} -\frac{m_1^2 m_2^2}{4} \right )\nonumber \\
& \phantom{={}} - \frac{1}{12} \hbar^2 \left (5 + \frac{u}{m_2} + m_1 m_2 u + 8 m_1 u^2 + \frac{u^2}{m_2^2} - 2 m_2^2 u^2 + m_1^2 m_2^2 u^2 \right ) + \CO(\hbar^4, u^3) \, .
\end{align}
\end{subequations}
Exponentiating the nontrivial A-period we find for the mirror map
\begin{equation}
u(Q_u)= Q_u-m_1 Q_u^3+(2 m_2+\frac{m_2}{4} \hbar^2) Q_u^4+(-3+m_1^2-\hbar^2) Q_u^5+2 m_1 m_2 \hbar^2 Q_u^6 + \CO(\hbar^4, Q_u^7) \, .
\end{equation}
Plugging this into the B-period we can integrate the special geometry relation. After inserting the relations eq.~(\ref{eq:relationsof2}) we get for the free energy
\begin{subequations}
\begin{align}
F^{(0,0)} &= \left (Q_1+Q_2\right ) + \left (\frac{Q_1^2}{8}-2 Q_1 Q_2+\frac{Q_2^2}{8}+Q_2 Q_3\right ) + \left (\frac{Q_1^3}{27}+\frac{Q_2^3}{27}-2 Q_1 Q_2 Q_3\right ) \nonumber \\
& \phantom{={}} + \left (\frac{Q_1^4}{64}-\frac{Q_1^2 Q_2^2}{4}+\frac{Q_2^4}{64}+3 Q_1 Q_2^2 Q_3+\frac{Q_2^2 Q_3^2}{8}\right ) + \left (\frac{Q_1^5}{125}+\frac{Q_2^5}{125}-4 Q_1^2 Q_2^2 Q_3\right ) \, ,\\
F^{(1,0)} &= \left (-\frac{Q_1}{24}-\frac{Q_2}{24}\right ) + \left (-\frac{Q_1^2}{48}-\frac{Q_1 Q_2}{6}-\frac{Q_2^2}{48}-\frac{Q_2 Q_3}{24}\right ) + \left (-\frac{Q_1^3}{72}-\frac{Q_2^3}{72}-\frac{Q_1 Q_2 Q_3}{6}\right ) \nonumber \\
& \phantom{={}} + \left (-\frac{Q_1^4}{96}-\frac{Q_1^2 Q_2^2}{12}-\frac{Q_2^4}{96}+\frac{7}{8} Q_1 Q_2^2 Q_3-\frac{Q_2^2 Q_3^2}{48}\right ) + \left (-\frac{Q_1^5}{120}-\frac{Q_2^5}{120}-\frac{7}{3} Q_1^2 Q_2^2 Q_3\right ) \, .
\end{align}
\end{subequations}
As for local $\IF_2$ (section \ref{sec:localF2}), we see a discrepancy with the computations in the $A$-model concerning
the instanton number $\hat{n}_n^{0,0,1} = 1$
\subsection{\texorpdfstring{A mass deformation of the local $E_8$ del Pezzo}{A mass deformation of the local E8 del Pezzo}}
\label{sec:localE8}
\begin{figure}[h]
 \begin{center}
  \includegraphics[scale=0.2]{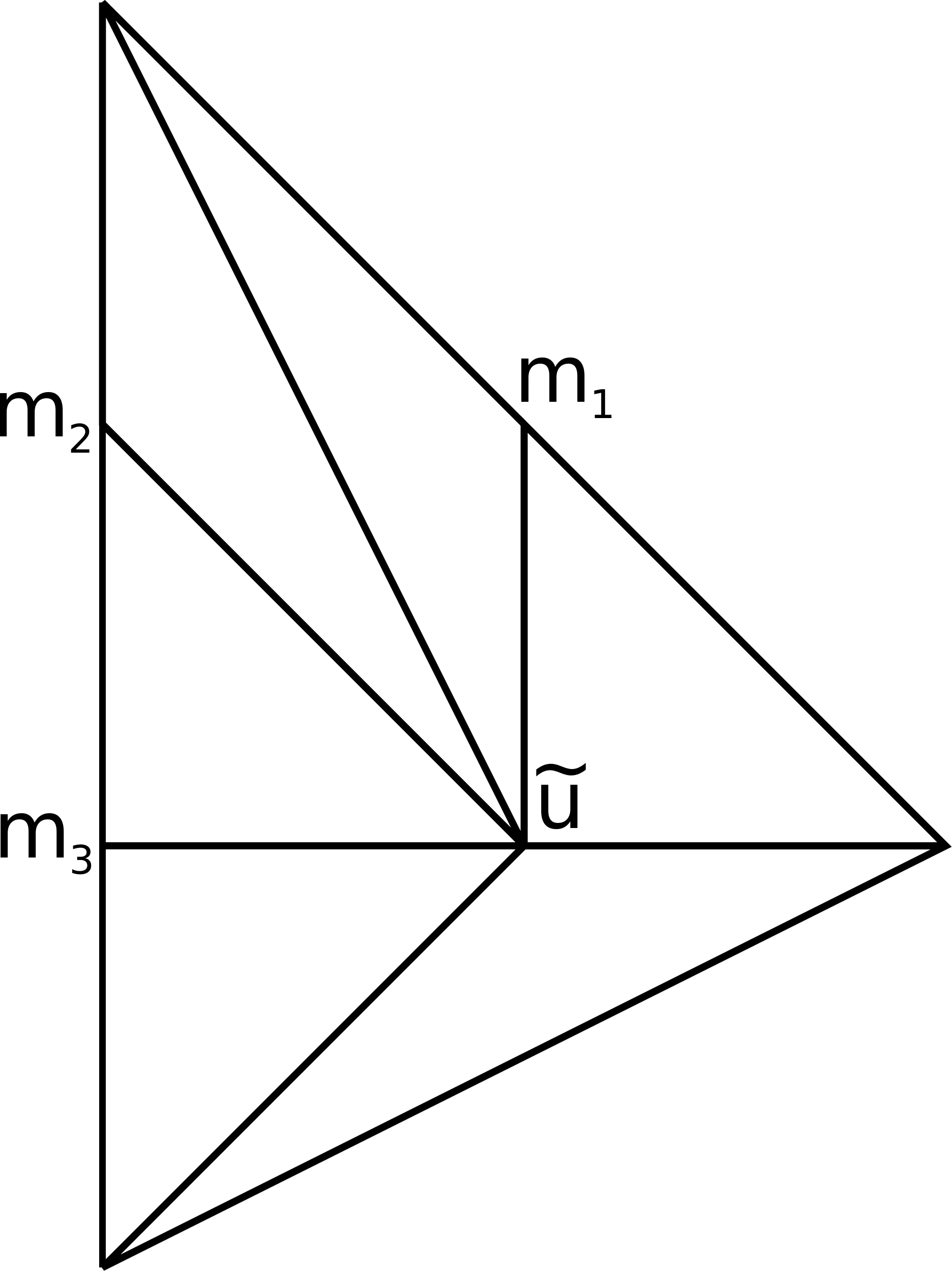}
  \caption{Toric diagram of the mass deformed local $E_8$ de Pezzo with the assigned masses.}
  \label{fig:e8}
  \end{center}
\end{figure}
\begin{equation} 
 \label{datae8} 
 \begin{array}{cc|crr|rrrrl|} 
    \multicolumn{5}{c}{\nu_i }    &l^{(1)}&l^{(2)}&l^{(3)}&l^{(4)}&\\ 
    D_u    &&     1&     0&   0&           0&-1& 0& 0&\\ 
    D_1    &&     1&     1&   0&           1& 0& 0& 0&\\ 
    D_2    &&     1&     0&   1&          -2& 1& 0& 0&\\ 
    D_3    &&     1&    -1&   2&           1&-1& 1& 0&\\ 
    D_4    &&     1&    -1&   1&           0& 1&-2& 1&\\ 
    D_5    &&     1&    -1&   0&           0& 0& 1&-2&\\ 
    D_6    &&     1&    -1&   -1&          0& 0& 0& 1& \\ 
  \end{array} \ . 
\end{equation} 
This geometry is denoted in \cite{Huang:2013yta} as polyhedron $10$. The diagram can be found in fig~(\ref{fig:e8}). The toric data is given in eq.~(\ref{datae8}). With the toric data we find for the coordinates
\begin{equation}
z_1=\frac{1}{m_1^2}\;,\quad z_2=\frac{m_1 m_2}{\tilde{u}}\;,\quad z_3=\frac{m_3}{m_2^2}\;,\quad z_4=\frac{m_2}{m_3^2}\, .
\end{equation}
Defining $u=\frac{1}{\tilde{u}}$ we find for the quantum mirror curve in the $u$ and $m$ coordinates
\begin{equation}
	H=1+e^x+e^p+m_3 u^2 e^{2p} - \frac{m_1 m_3}{m_2} u e^{\frac{\hbar}{2}+p+x}+\frac{m_3}{m_2^2} e^{2x}+\frac{m_2}{m_3^2} e^{-x} \, .
\end{equation}
The coefficients of the classical Weierstrass normal form are
\begin{subequations}
\begin{align}
g_2(u, m_1, m_2, m_3) &= 27 u^4 (1 - 8 m_3 u^2 + 24 m_1 u^3 - 48 m_2 u^4 + 16 m_3^2 u^4) \, ,\\
g_3(u, m_1, m_2, m_3) &= 27 u^6 (1 - 12 m_3 u^2 + 36 m_1 u^3 - 72 m_2 u^4 + 48 m_3^2 u^4 - 144 m_1 m_3 u^5 \nonumber \\
& \phantom{={}} - 864 u^6 + 216 m_1^2 u^6 + 288 m_2 m_3 u^6 - 64 m_3^3 u^6) \, .
\end{align}
\end{subequations}
The resulting Schr\"odinger equation can be solved perturbatively in $\hbar$ and gives for the zeroth order WKB function
\begin{equation}
  S_0^\prime (x)= \log \left(\frac{-1+e^{x} z_2 z_3-e^{-x}\sqrt{e^x (e^x (-1+e^x z_2 z_3)^2-4 z_1 z_2^2 z_3 (e^x+e^{2 x}+e^{3 x} z_3+z_4))}}{2 z_1 z_2^2 z_3}\right ) \, .
\end{equation}
The second order WKB function can be calculated by use of the following operator up to exact terms out of the zeroth order
\begin{align}
  \CDD_2 &  =\frac{1}{6 \delta} \left(  u^2 \left(4 m_3^2 u^2 \left(m_2 m_1 u+m_1^2-6\right)+m_3 \left(-9 m_1 \left(m_1^2-4\right) u^3+4 m_2^2 u^2-5 m_1 m_2 u+6\right) \right.\right. \nonumber \\
   & \phantom{ =\frac{1}{6 \delta} } + \left.\left. 6 u \left(2 m_1 m_2 u-3\right) \left(m_1-2 m_2 u\right)\right) \right)
  \Theta_u \nonumber \\
 & \phantom{={}} +  \frac{1}{24 \delta}  \left(4 m_1 \left(4 m_2 \left(m_3^2-6 m_2\right)-9 \left(m_1^2-4\right) m_3\right) u^5+4 \left(4 m_3 m_2^2+3 \left(5 m_1^2+12\right) m_2 +  \right.\right. \nonumber \\
 & \phantom{= +  \frac{1}{24 \delta}}  \left.\left. +4 \left(m_1^2-6\right) m_3^2\right) u^4
 +3 m_1 \left(m_1^2-8 m_2 m_3-36\right) u^3 \right. \nonumber \\
 & \phantom{= +  \frac{1}{24 \delta}}  \left. -4 \left(m_2^2+\left(m_1^2-12\right) m_3\right) u^2+5 m_1 m_2 u-6\right)
  \Theta_u^2 
  \label{eqn:E8_HOp}
\end{align}
where
\begin{equation}
 \delta = m_1 \left(9 m_1^2+4 m_2 m_3-36\right) u^3-8 \left(m_2^2+\left(m_1^2-3\right) m_3\right) u^2+7 m_1 m_2 u-6\,.
\end{equation}
With this we find for the quantum corrected nontrivial periods
\begin{subequations}
\begin{align}
a&= \log(u) + m_3 u^2 - 2 m_1 u^3 + 3 m_2 u^4 + \frac{3 m_3^2 u^4}{2} \nonumber \\
&\phantom{= \log(u) + m_3 u^2 - 2 m_1 u^3} {} + \left(-\frac{5 m_1 u^3}{4} + m_2 u^4 + \frac{1}{2} m_1^2 m_2 u^4\right) \hbar^2 + \CO(\hbar^4, u^5)\, ,\\
a_D&= - 3 \log(u)^2 - 6 m_3 u^2 \log(u) + m_1 m_2 u - \frac{m_2^2 u^2}{2} + \frac{1}{4} m_1^2 m_2^2 u^2 - 3 m_3 u^2 - \frac{1}{2} m_1^2 m_3 u^2 \nonumber \\
&+ \left (-\frac{1}{4} + \frac{m_1 m_2 u}{8} + \frac{m_2^2 u^2}{12} - \frac{5}{72} m_1^2 m_2^2 u^2 - \frac{m_3 u^2}{2} + \frac{1}{12} m_1^2 m_3 u^2 \right ) \hbar^2 + \CO(\hbar^4, u^3)\, .
\end{align}
\end{subequations}
This leads to the following mirror map after exponentiating the nontrivial A-period
\begin{align}
u(Q_u)&= Q_u-m_3 Q_u^3+2 m_1 Q_u^4-3 m_2 Q_u^5+m_3^2 Q_u^5+\frac{1}{12} (15 m_1 Q_u^4-12 m_2 Q_u^5\nonumber \\
&-6 m_1^2 m_2 Q_u^5-12 m_1 m_2^2 Q_u^6+7 m_1^3 m_2^2 Q_u^6+72 m_1 m_3 Q_u^6-12 m_1^3 m_3 Q_u^6) \hbar^2 + \CO(\hbar^4, Q_u^7)\, .
\end{align}
Now we can integrate the special geometry relation and plug in 
the following relations between the mass parameters and the coordinates
\begin{subequations}
\begin{align}
m_1&= \frac{1+Q_1}{\sqrt{Q_1}}\, ,\\
m_2&= \frac{1+Q_3+Q_3 Q_4}{Q_3^{2/3} Q_4^{1/3}}\, ,\\
m_3&= \frac{1+Q_4+Q_3 Q_4}{Q_3^{1/3}Q_4^{2/3}} \, .
\end{align}
\end{subequations}
These relations do not get any quantum corrections as was expected since the mass parameters $m_i$ are trivial parameters. 
Using additionally $Q_t=Q_1^{\frac{1}{2}}Q_2 Q_3^{\frac{2}{3}} Q_4^{\frac{1}{3}}$ we find for the refined free energies in the Nekrasov-Shatashvili limit
\begin{subequations}
\begin{align}
 \CW_0 & =
\Li_3^{{(0,1,0,0)}}+\Li_3^{{(0,1,1,0)}}+\Li_3^{{(0,1,1,1)}}+\Li_3^{{(1,1,0,0)}}+\Li_3^{{(1,1,1,0)}}+\Li_3^{{(1,1,1,1)}}-2 \Li_3^{{(1,2,1,0)}} \nonumber \\
& \phantom{={}} {} -2 \Li_3^{{(1,2,1,1)}} -2 \Li_3^{{(1,2,2,1)}}+3 \Li_3^{{(1,3,2,1)}}+3 \Li_3^{{(2,3,2,1)}}-4 \Li_3^{{(2,4,2,1)}}-4 \Li_3^{{(2,4,3,1)}} \nonumber \\
& \phantom{={}} {} -4 \Li_3^{{(2,4,3,2)}}+5 \Li_3^{{(2,5,3,1)}}+5 \Li_3^{{(2,5,3,2)}}+5 \Li_3^{{(3,5,3,1)}} \\
  -24 \CW_1 & =\Li_1^{{(0,1,0,0)}}+\Li_1^{{(0,1,1,0)}}+\Li_1^{{(0,1,1,1)}}+\Li_1^{{(1,1,0,0)}}+\Li_1^{{(1,1,1,0)}}+\Li_1^{{(1,1,1,1)}}+4 \Li_1^{{(1,2,1,0)}} \nonumber \\
 & \phantom{={}} {}  +4 \Li_1^{{(1,2,1,1)}}+4 \Li_1^{{(1,2,2,1)}}-21 \Li_1^{{(1,3,2,1)}}-21 \Li_1^{{(2,3,2,1)}}+56 \Li_1^{{(2,4,2,1)}}+56 \Li_1^{{(2,4,3,1)}} \nonumber \\
 & \phantom{={}} {} +56 \Li_1^{{(2,4,3,2)}}-115 \Li_1^{{(2,5,3,1)}}-115 \Li_1^{{(2,5,3,2)}}-115 \Li_1^{{(3,5,3,1)}}
   \, .
\end{align}
\end{subequations}
Here we defined $\Li_n^{(\beta)} = \Li_n(Q^\beta)$.
\section{Conclusions}
By quantizing the special geometry of local Calabi-Yau 
manifolds related to the del Pezzo surfaces we solved the 
topological string in the Nekrasov-Shatashvili limit 
for many new geometries. We confirmed the quantization approach 
in the large radius limit for $\mathbb{F}_0$, $\mathbb{F}_1$, 
$\mathbb{F}_2$, as well as for the blown up surfaces  
${\cal B}_2(\mathbb{P}^2)$ and ${\cal B}_1(\mathbb{F}_2)$ and 
a mass deformed $E_8$ del Pezzo surface. 

The mass deformation parameters $m_i$ and the modular Coulomb branch 
parameter $u$, also called non-normalizable moduli and 
normalizable moduli are clearly distinguished in our approach. 
For the relevant genus one mirror curves the structure is encoded 
in a third order differential operator in the modular parameter 
with rational coefficients in the $m_i$ determining the two 
classical periods $a(u,m_i)$ and $a_D(u,m_i)$. These two periods are the only objects that get quantum deformed. The
quantum deformed periods are defined by applying one 
differential operator ${\cal D}^{(2)}(u, m_i; \hbar)$ 
to the classical periods. This operator is second order in the 
modular parameter with rational coefficients in the mass 
parameters, but so far we could only determine it perturbatively  
in $\hbar$. However given ${\cal D}^{(2)}(u,  m_i; \hbar)$ to 
some order in $\hbar$ we can immediatly determine the quantum deformation
perturbativley at any point in the $(u,m_i)$ space. 
With this information we can predict and in some cases check the 
orbifold  and conifold expansions for the quantum deformed free 
energy.

We only considered the closed sector though and it would be very interesting
to see whether the wavefunctions which solve the Schr\"odinger equations 
also compute correct open amplitudes or if they are only useful for
deriving quantum deformed meromorphic differentials which are evaluated
over closed contours.

The way the quantum special geometry was derived somewhat suggested that
we take the zeroth order contributions to the periods and deform them by
a parameter $\hbar$. Considering that the Picard-Fuchs operators
annihilate the zeroth order contributions,  maybe also a
Picard-Fuchs operator that annihilates the quantum deformed periods
exists.

We also used the difference equation ansatz to derive the free energies
of local $\IF_0$ and local $\IF_1$ at large radius. For the conifold
and orbifold point however, we were not able to extract the 
necessary data to solve the problem in this way. This computation
would lead to an expression exact in $\hbar$, which is an 
expression we do not yet have for the topological string B-model.

One problem we encountered are certain missing instanton numbers 
corresponding to K\"ahler parameters related to non-normalizable divisors.
These are not captured by the Picard-Fuchs system, like e.\,g. in the case
of the resolved conifold. We still were able to apply our methods by 
making use of \cite{Forbes:2005xt}, where it was noted, that we can find
the generating series for the B-cycle via the Frobenius-method. 

The Schr\"odinger equation for brane-wavefunctions in the full refined topological string
depends on multiple times, which are the K\"ahler parameters. Having our results in mind, 
it would certainly be important to carefully distinguish between
normalizable and non-normalizable moduli when analyzing this problem in full
generality.
\section{Acknowledgements}
The authors would like to thank Daniel Krefl and Hans Jockers for helpful diskussions. 
MH is supported by the "Young Thousand People" plan by the Central Organization Department in China. MH thanks University of Bonn, and University of Wisconsin, Madison for hospitality during parts of the work.
AK and MS are supportted by the DFG grant KL2271/1-1.
The work of JR is supported by a scholarship of the graduate school BCGS.
%
%
%
%
%
\newpage
\appendix
\section{Eisenstein series}\label{sec:eisenstein}
The divisor function $\sigma_x$ is defined by
\begin{equation}
  \sigma_x(n) = \sum_{d | n} d^x
\end{equation}
and the Eisenstein series $E_4$ and $E_6$ are defined by
\begin{align}
  E_4(\tau) & = 1 + 240 \sum_{n=1}^\infty \sigma_3 (n) q^n \\
  E_6(\tau) & = 1 -504 \sum_{n=1}^\infty \sigma_5 (n) q^n
\end{align}
in terms of it. The parameter $q$ is defined by 
\begin{equation}
  q = e^{2 \pi \ri \tau}
\end{equation}
\section{\texorpdfstring{local $\IF_0$}{local F0} } \label{app:localF0}
The higher order operators are
\begin{align}
\CDD_2 & = \frac{1}{6} (-u - m u) \Theta_u + \frac{1}{12} (1 - 4 u - 4 m u)\Theta_u^2 \, ,\\
\CDD_4&= -\frac{1}{180 \Delta^2}u^2 (64 m^5 u^3 + (-1 + 4 u)^3 - 48 m^4 u^2 (1 + 116 u) + 4 m^3 u (3 + 328 u + 1376 u^2) \nonumber \\
  & \phantom{={}} - 
    4 m (8 - 37 u - 328 u^2 + 1392 u^3) + 
    m^2 (-1 + 148 u - 6112 u^2 + 5504 u^3)) \Theta_u \nonumber \\
& \phantom{={}} -\frac{1}{360 \Delta^2}
u ((1 - 4 u)^4 + 256 m^5 u^4 - 256 m^4 u^3 (1 + 87 u) + 32 m^3 u^2 (3 + 214 u + 688 u^2) \nonumber \\
& \phantom{={}} - m (1 - 4 u)^2 (-1 + 268 u + 1392 u^2) + 
    16 m^2 u (-1 + 48 u - 1720 u^2 + 1376 u^3))\Theta_u^2 \, ,
\end{align}
with $\Delta =(1 - 8 (1 + m) u + 16 (-1 + m)^2 u^2)$.
And some higher WKB functions are
\begin{align}
  S'_0 (x) & = \log \left(\frac{1}{2} e^{-x} (e^x-e^{2 x}-z_1+\sqrt{(-e^x+e^{2 x}+z_1)^2-4 e^{2 x} z_2})\right)\, ,\\
  S'_1 (x) & = \frac{e^{3 x}-e^{4 x}-e^x z_1+z_1^2}{2 (-2 e^{3 x}+e^{4 x}-2 e^x z_1+z_1^2+e^{2 x} (1+2 z_1-4 z_2))} \, ,\\
  S'_2 (x) & =-\frac{1}{12 ((-e^x+e^{2 x}+z_1)^2-4 e^{2 x} z_2)^(5/2)} e^x (e^{8 x}+z_1^4-e^x z_1^3 (3+4 z_1-22 z_2) \nonumber \\
  & \phantom{={}} +e^{6 x} (3+16 z_1-18 z_2)  +e^{2 x} z_1^2 (3+16 z_1-18 z_2)+e^{7 x} (-3-4 z_1+22 z_2) \nonumber \\
  & \phantom{={}} +2 e^{4 x} z_1 (5+15 z_1+34 z_2) -e^{5 x} (1+12 z_1^2+4 z_2-32 z_2^2+z_1 (21+38 z_2)) \nonumber \\
&\phantom{={}} -e^{3 x} z_1 (1+12 z_1^2+4 z_2-32 z_2^2+z_1 (21+38 z_2))) \, .
\end{align}
\subsection{Orbifold point}\label{app:F0_orbifold_point}
Here we present the raw data of the computation at the orbifold point
in terms of periods $s_t$ and $s_m$ without having fixed the constants
of integration. Also neither the shift in $\hbar$ nor the normalization
have been carried out.
\begin{subequations}
\begin{align}
 \tilde{F}^{(0,0)} & = c_0(s_m) + f_0^6 +
 \frac{\log{s_m} s_t^2}{2}+\frac{s_t^2}{2} +
 \frac{s_t^4}{1152}-\frac{1}{384} s_m^2 s_t^2
 -\frac{31 s_m^2 s_t^4}{1474560} \nonumber \\
&\phantom{=} +\frac{73 s_m^4 s_t^2}{2949120} +\frac{283 s_t^6}{44236800} + \CO(s_i^7) \\
 \tilde{F}^{(1,0)} & = c_1(s_m) + f_1^6 + \frac{7}{1152} s_2^2
  -\frac{253}{1\ts{}474\ts{}560} s_1^2 s_2^2  
  + \frac{511}{4\ts{}423\ts{}680} s_2^4 
    \nonumber \\
  & \phantom{{} = {}} +  \frac{2959}{594\ts{}542\ts{}592} s_1^4 s_2^2  
  -  \frac{1103}{148\ts{}635\ts{}648} s_1^2 s_2^4 
  + \frac{29\ts{}923}{8\ts{}918\ts{}138\ts{}880} s_2^6  + \CO(s_i^7) \\
  \tilde{F}^{(2,0)} & =
   c_2(s_m) + f_2^6 + \frac{9631}{44\ts{}236\ts{}800}  s_2^2  
  -\frac{8089}{424\ts{}673\ts{}280}s_1^2 s_2^2 + \frac{1489}{79\ts{}626\ts{}240} s_2^4  + 
  \nonumber\\
  &  \phantom{{} = {}} + \frac{9\ts{}712\ts{}951}{8\ts{}697\ts{}308\ts{}774\ts{}400} s_1^4 s_2^2  - \frac{10\ts{}152\ts{}757}{4\ts{}348\ts{}654\ts{}387\ts{}200} s_1^2 s_2^4 + 
   \frac{5\ts{}466\ts{}903\ts{}857}{260\ts{}919\ts{}263\ts{}232\ts{}000}s_2^6  + \CO(s_i^7) \\
  \tilde{F}^{(3,0)} & = c_3(s_m) +  f_3^6 + 
  \frac{1146853 s_t^2}{62426972160} +
  \frac{373588141 s_t^4}{91321742131200}-\frac{98735143 s_m^2 s_t^2}{30440580710400} 
  \nonumber\\
  &  \phantom{{} = {}} 
   -\frac{170286827 s_m^2 s_t^4}{200907832688640}+\frac{1031514229 s_m^4 s_t^2}{3214525323018240}+\frac{28374740293 s_t^6}{48217879845273600}
   + \CO(s_i^7)
\end{align}
\end{subequations}
where
\begin{align}
 f_0^n&  = - \sum_{i=1}^n \frac{1}{4 i \left(2 i^2+3 i+1\right)} \frac{s_f^{2 i + 2}}{s_m^{2 i}} \\
 f_1^n&  = \sum_{i=1}^n \frac{1}{12 i} \frac{s_f^{2 i}}{s_m^{2 i}} \\
 f_2^n&  = - \sum_{i=1}^n \frac{7 (2 i+1)}{360 } \frac{s_f^{2 i}}{s_m^{2 i+2}} \\
 f_3^n&  = \sum_{i=1}^n \frac{31 \left(4 i^3+12 i^2+11 i+3\right)}{7560} \frac{s_f^{2 i}}{s_m^{2 n+4}} \, .
\end{align}
There are some additional terms of order zero in $s_f/s_m$, which are suppressed
if we go to higher orders in the expansion, hence we dropped them here.
\section{\texorpdfstring{$\CO(-3) \rightarrow \IP^2$}{O(-3) -> P2} }  \label{app:localP2}
The A-periods:
\begin{align}
 a^{(0)} & = \log (u) -2 u^3 +15 u^6 -\frac{560 u^9}{3}+ \frac{5775 u^{12}}{2} + \CO(u^{15}) \\
 a^{(2)} & = -\frac{u^3}{4} +\frac{15 u^6}{2} -210 u^9 + 5775 u^{12}  + \CO(u^{15})  \\
 a^{(4)} & = -\frac{u^3}{192} +\frac{13 u^6}{8} -\frac{987 u^9}{8} + 6545 u^{12}  + \CO(u^{15}) 
\end{align}
The mirror map:
\begin{align}
 u|_{\hbar^0} & = Q_t + 2 Q_t^4 -Q_t^7 + 20 Q_t^{10} + \CO(Q_t^{13} ) \\
 u|_{\hbar^2} & = \frac{Q_t^4}{4}-4 Q_t^7 + \frac{145 Q_t^{10}}{2} + \CO(Q_t^{13} ) \\
 u|_{\hbar^4} & =  +\frac{Q_t^4}{192} -\frac{4 Q_t^7}{3} + \frac{7549 Q_t^{10}}{96}+ \CO(Q_t^{13}) 
\end{align}
The B-periods:
\begin{align}
  a^{(0)}_D & = 9 u^3 -\frac{423 u^6}{4} +1486 u^9 -\frac{389415 u^{12}}{16} + \CO(u^{13})   \\
  a^{(2)}_D & = -\frac{1}{8} +\frac{21 u^3}{8} -\frac{603 u^6}{8} +\frac{8367 u^9}{4} -\frac{458715 u^{12}}{8}  + \CO(u^{13}) \\
  a^{(4)}_D & = \frac{87 u^3}{640}  -\frac{3633 u^6}{160} +\frac{485649 u^9}{320} -\frac{607657 u^{12}}{8} + \CO(u^{13})
\end{align}

\subsection{Orbifold point}
\label{app:P2_orbifold}
The periods $\sigma$ are given by
\begin{subequations}
 \label{eqn:P2_orb_a} 
\begin{align}
 (-1)^{2/3} \sigma^{(0)} & = 
 -3 \psi - \frac{1}{8}\psi^4 - \frac{4}{105} \psi^7 - \frac{49}{ 2700 } \psi^{10} - \frac{245}{ 23\ts{}166 } \psi^{13} + \CO( \psi^{16} )\\
 (-1)^{2/3} \sigma^{(2)} & = -\frac{1}{24} \psi - \frac{1}{36}\psi^4  - \frac{7}{ 270 } \psi^7 - \frac{49}{ 1944} \psi^{10} - \frac{3185}{128\ts{}304} \psi^{13} + \CO( \psi^{16} )\\
 (-1)^{2/3}\sigma^{(4)} & = - \frac{23}{ 17\ts{}280 } \psi - \frac{11}{1620} \psi^4 - \frac{637}{38\ts{}880}\psi^7 - \frac{2107}{69\ts{}984} \psi^{10} - \frac{886\ts{}067}{18\ts{}475\ts{}776} \psi^{13} + \CO( \psi^{16} ),  
\end{align}
\end{subequations}
while the dual periods are given by
\begin{subequations}
\label{eqn:P2_orb_aD}
\begin{align}
 (-1)^{1/3}\sigma_D^{(0)} & = - \frac{3}{2} \psi^2  - \frac{1}{5} \psi^5  - \frac{25}{336} \psi^8 - \frac{80}{2079} \psi^{11} -  \frac{1210}{51\ts{}597} \psi^{14} + \CO(\psi^{17})
   \\
 (-1)^{1/3}\sigma_D^{(2)} & = -\frac{1}{12} \psi^2  - \frac{5}{72} \psi^5 - \frac{25}{378} \psi^8 - \frac{110}{1701} \psi^{11} - \frac{605}{9477} \psi^{14} + \CO(\psi^{17})\\
 (-1)^{1/3}\sigma_D^{(4)} & = -\frac{1}{144} \psi^2 - \frac{85}{3456} \psi^5 - \frac{10}{189} \psi^8 - \frac{3751}{40\ts{}824} \psi^{11} - \frac{16\ts{}093}{113\ts{}724} \psi^{14} + \CO(\psi^{17}) . 
\end{align}
\end{subequations}
\subsection{Conifold point}
\label{app:P2_conifold}
\begin{subequations}
\label{eqn:P2_con_a}
\begin{align}
t^{(0)} & = \Delta +\frac{11\, \Delta^2}{18}+ \frac{109\, \Delta^3}{243} +\frac{9389\, \Delta^4}{26244} 
+\frac{88351\, \Delta^5}{295245} + \CO( \Delta^6)\\
t^{(2)} & = \frac{1}{36} + \frac{\Delta}{324} + \frac{5\, \Delta^2}{4374}
+\frac{35\, \Delta^3}{59049} + \frac{385\, \Delta^4}{1062882} 
+ \frac{7007\, \Delta^5}{28697814} + \CO( \Delta^6 ) \\
t^{(4)} & = \frac{19}{139968} -\frac{91\, \Delta}{1259712} -\frac{89\, \Delta^2}{2834352}
-\frac{3521\, \Delta^3}{229582512} -\frac{34265\, \Delta^4}{4132485216}
-\frac{179179\, \Delta^5}{37192366944} + \CO( \Delta^6 )\,.
\end{align}
\end{subequations}
The first corrections to the dual period are
\begin{subequations}
\label{eqn:P2_con_aD}
\begin{align}
{t^{(0)}_c}_D & =  \text{a0}
+ a_1 t^{(0)}_c - \frac{1}{2 \pi \ri}\left(
t_c^{(0)} \log(\Delta) + 
\frac{7 \Delta^2}{12} +\frac{877 \Delta^3}{1458} +\frac{176015 \Delta^4}{314928}
+\frac{9065753 \Delta^5}{17714700} + \CO(\Delta^6)
 \right) \\
{t^{(2)}_c}_D & =  \phantom{ \text{a0} + {} }
 a_1 t^{(2)}_c - \frac{1}{2 \pi \ri}\left(
 t_c^{(2)} \log(\Delta) + 
\frac{1}{8 \Delta} +\frac{\Delta}{108} +\frac{211 \Delta^2}{52488} +\frac{3139 \Delta^3}{1417176}
+\frac{35663 \Delta^4}{25509168}+ \CO(\Delta^5)
 \right) \\
{t^{(4)}_c}_D & =  \phantom{ \text{a0} + {} }
 a_1 t^{(4)}_c \nonumber \\
&\phantom{=} {}- \frac{1}{2 \pi \ri}\left(
 t_c^{(4)} \log(\Delta) +
 \frac{7}{320 \Delta^3}-\frac{251}{5760 \Delta^2}+\frac{247}{10368 \Delta}
 -\frac{691}{419904}  - \frac{941 \Delta}{3779136}
+ \CO(\Delta^2)
 \right)
\end{align}
\end{subequations}
In these expressions we use
\begin{equation}
 a_0=-\frac{\pi}{3} -1.678699904 \ri=
\frac{1}{i\sqrt{3}\Gamma\left(\frac{1}{3}\right)\Gamma\left(\frac{2}{3}\right)}G^{3\,3}_{2\,2}
\left({{\frac{1}{3}\ \frac{2}{3} \ 1} \atop {0 \ 0 \ 0}}\biggr| -1\right)
\quad\text{and}\quad
a_1 = \frac{3 \log(3)+1}{2\pi \ri}\,.
\end{equation}
\section{\texorpdfstring{local $\IF_1$}{local F1} } \label{app:localF1}
The higher order operators are
\begin{align}
\CDD_2 &= \frac{m u^2\left(4m-9u\right)}{6\left(-8m+9u\right)}\Theta_u + \frac{4m-3u-16m^2 u^2+36u^3 m}{24(8m-9u)}\Theta_u^2 \, , \\
\CDD_4&= -\frac{1}{2880 (8 m - 9 u) (m - 
      u - 8 m^2 u^2 + 36 m u^3 - 27 u^4 + 16 m^3 u^4)^2}
u^3 ( 
      27 u^3 (-5 + 999 u^3) \nonumber \\
&\phantom{={}} + 1536 m^7 u^5 (-2 + 121 u^3)  + 768 m^6 (u^3 + 209 u^6)+ 2 m^3 (53 - 23148 u^3 + 8856 u^6) \nonumber \\
& \phantom{={}} - 2 m^2 u (163 - 49113 u^3 + 52488 u^6) + 
      16 m^4 u^2 (457 + 8802 u^3 + 58806 u^6) \nonumber \\
& \phantom{={}} +4096 m^8 u^7 - 64 m^5 (u + 1292 u^4 + 12582 u^7) + 
      m (453 u^2 - 81972 u^5 + 87480 u^8))\Theta_u \nonumber \\
& \phantom{={}} -\frac{1}{5760 (8 m - 9 u) (m - u - 8 m^2 u^2 + 36 m u^3 - 27 u^4 + 16 m^3 u^4)^2}
u (8192 m^8 u^9 \nonumber \\
& \phantom{={}} + 1024 m^7 u^7 (-8 + 363 u^3) + 
      768 m^6 u^5 (4 + 401 u^3) + 27 u^5 (-29 + 2619 u^3) \nonumber \\
& \phantom{={}} - 36 m^3 u^2 (-15 + 4060 u^3 + 1416 u^6) - 
      128 m^5 u^3 (4 + 1673 u^3 + 12168 u^6) \nonumber \\
& \phantom{={}} + m^2 (7 - 1780 u^3 + 289260 u^6 - 136080 u^9) + 
      m u (7 + 2796 u^3 - 222264 u^6 + 174960 u^9) \nonumber \\
& \phantom{={}} + 32 m^4 (u + 819 u^4 + 12114 u^7 + 58806 u^10))\Theta_u^2
\end{align}

\begin{align}
  S'_0 (x) & = \log \left(\frac{1}{2} e^{-x} \left(e^x-e^{2 x}-z_1 -\sqrt{\left(-e^x+e^{2 x}+z_1\right)^2+4 e^{3 x} z_2}  \right)\right) \, ,\\
  S'_1 (x) & = -\frac{e^x z_1+e^{3 x} \left(2 z_2-1\right)+e^{4 x}-z_1^2}{2 \left(\left(e^x \left(e^x-1\right)+z_1\right){}^2+4 e^{3 x} z_2\right)} \, ,\\
  S'_2(x) & = -\frac{5 \left(e^x+e^{2 x}-3 z_1\right){}^2 \left(e^x \left(e^x-1\right)+z_1\right){}^3}{32 \left(\left(-e^x+e^{2 x}+z_1\right){}^2+4 e^{3 x} z_2\right){}^{5/2}}  \nonumber\\
  	& \phantom{={}} + \frac{\left(e^x \left(e^x-1\right)+z_1\right) \left(-6 e^x \left(3 e^x+23\right) z_1+e^{2 x} \left(e^x \left(19 e^x+14\right)+19\right)+171 z_1^2\right)}{96 \left(\left(-e^x+e^{2 x}+z_1\right){}^2+4 e^{3 x} z_2\right){}^{3/2}}\nonumber \\
  	&  \phantom{={}} -\frac{e^x \left(e^x-1\right)+9 z_1}{24 \sqrt{\left(-e^x+e^{2 x}+z_1\right){}^2+4 e^{3 x} z_2}}\, .
\end{align}
\section{\texorpdfstring{local $\IF_2$}{local F2} } \label{app:localF2}
Some higher WKB functions are
\begin{align}
    S_0^\prime (x)&= \log \left (\frac{1}{2} \left (-1-e^x-e^{2 x} m u -\sqrt{(1+e^x+e^{2 x} m u^2)^2-\frac{4}{m^2}}\right ) \right ) \\
S_1^\prime (x)&= -\frac{e^x m^2 (1 + 3 e^{2 x} m u^2 + 2 e^{3 x} m^2 u^4 + e^x (1 + 2 m u^2))}{2 (-4 + m^2 (1 + e^x + e^{2 x} m u^2)^2)}\\
S_2^\prime (x)&= \frac{1}{12 (-4 + m^2 (1 + e^x + e^{2 x} m u^2)^2)^{5/2}} e^x m^5 \left (1 - \frac{32}{m^4} + \frac{4}{m^2} + e^{8 x} m^4 u^8 \right . \nonumber \\
  & \phantom{={}} {} + e^{7 x} m^3 u^6 (3 + 4 m u^2) + e^{6 x} m^2 u^4 (3 + 16 m u^2) + 2 e^{4 x} u^2 (5 m - 86 u^2 + 15 m^2 u^2) \nonumber \\
  & \phantom{={}} {} + e^{2 x} (3 - \frac{22}{m^2} + \frac{16 (-4 + m^2) u^2}{m}) + e^x (3 - \frac{18}{m^2} + \frac{4 (-32 + 4 m^2 + m^4) u^2}{m^3}) \nonumber \\
  & \phantom{={}} {} \left . + e^{5 x} m u^2 (1 + 21 m u^2 - 88 u^4 + 12 m^2 u^4) + e^{3 x} (1 + (21 - \frac{106}{m^2}) m u^2 + 12 (-6 + m^2) u^4) \right ) \, .
\end{align}
Additional operators are
\begin{align}
  \CDD_2 &=-\frac{1}{6} (m u) \Theta_u + \frac{1}{12} (1-4 m u)\Theta_u^2 \, ,\\
  \CDD_4 &=\frac{u^2}{180 \Delta^2}  \left(-4 m \left(3 m^2+28\right) u+m^2-64 m \left(m^4-92 m^2+352\right) u^3 \right. \nonumber \\
    & \phantom{=\frac{u^2}{180 \Delta^2}  ( } {} \left. +16 \left(3 m^4-94 m^2+552\right) u^2+30\right) \Theta_u \nonumber\\
    & \phantom{={}} {}+ \frac{u}{360 \Delta^2} \left(-96 m \left(m^2+5\right) u^2+4 \left(4 m^2+61\right) u-256 m \left(m^4-92 m^2+352\right) u^4 \right. \nonumber \\
    &\phantom{=+ \frac{u}{360 \Delta^2} ( } {} \left. +64 \left(4 m^4-123 m^2+652\right) u^3-m\right) \Theta_u^2
\end{align}
\begin{align}
\CDD_6 &= \frac{1}{7560 \Delta^4} u^4 \left (45056 m^8 u^{6} - 32768 m^9 u^{7} - 
    2048 m^7 u^{5} (12 + 4883 u^2) \right .\nonumber\\
  & \phantom{= \frac{1}{7560 \Delta^4} {}} {} + 1280 m^6 u^4 (5 + 5168 u^2) + 128 m^5 u^3 (-5 + 542 u^2 + 9792 u^4) \nonumber \\
  & \phantom{= \frac{1}{7560 \Delta^4} {}} - 16 m^4 u^2 (3 + 66744 u^2 + 8158976 u^4) \nonumber \\ 
  & \phantom{= \frac{1}{7560 \Delta^4} {}} {} - 32 m u (-32 + 58757 u^2 + 8863616 u^4 + 40161280 u^{6}) \nonumber \\
& \phantom{= \frac{1}{7560 \Delta^4} {}} {}
+ 
    8 m^3 u (2 + 40573 u^2 + 8116224 u^4 + 59797504 u^{6}) \nonumber \\
& \phantom{= \frac{1}{7560 \Delta^4} {}} {} + 6 (7 + 39032 u^2 + 7728128 u^4 + 123830272 u^{6}) \nonumber \\
& \phantom{= \frac{1}{7560 \Delta^4} {}} {} \left . + 
    m^2 (-1 - 36328 u^2 - 4319232 u^4 + 227704832 u^{6}) \right ) \Theta_u \nonumber\\
& \phantom{={}} {} + \frac{1}{15120 \Delta^4} u \left (212992 m^8 u^{7} - 131072 m^9 u^{8} - 
    4096 m^7 u^{6} (35 + 9766 u^2) \right .\nonumber \\
  & \phantom{=+ \frac{1}{15120 \Delta^4}{}}  + 1024 m^6 u^{5} (49 + 29990 u^2) + 
    256 m^5 u^4 (-35 - 5364 u^2 + 19584 u^4) \nonumber \\
  & \phantom{=+ \frac{1}{15120 \Delta^4}{}} - 
    64 m^4 u^3 (-7 + 81220 u^2 + 8528768 u^4) \nonumber \\
  & \phantom{=+ \frac{1}{15120 \Delta^4}{}} + 16 m^3 u^2 (7 + 120530 u^2 + 19097600 u^4 + 119595008 u^{6}) \nonumber \\
  & \phantom{=+ \frac{1}{15120 \Delta^4}{}}  + 
    4 m^2 u (-5 - 65394 u^2 - 6316032 u^4 + 221945856 u^{6}) \nonumber \\
  & \phantom{=+ \frac{1}{15120 \Delta^4}{}}  + 4 u (119 + 398200 u^2 + 59192320 u^4 + 790593536 u^{6}) \nonumber \\
  & \phantom{=+ \frac{1}{15120 \Delta^4}{}}  \left . + m (1 + 9320 u^2 - 10643584 u^4 - 1288617984 u^{6} - 
       5140643840 u^{8})\right) \Theta_u^2  \, ,
\end{align}
with $\Delta = (1 - 8 m u - 64 u^2 + 16 m^2 u^2)$.

\end{document}